\documentclass{ptephy_v1}%%%%where ptephy_v1 is the template name

\preprintnumber{XXXX-XXXX} %%% %%% Insert preprint number here
\usepackage{hyperref}
\usepackage{amsmath}% for dealing with mathematics,
\begin{document}

\title{Detection of the 4.4-MeV gamma rays from $^{16}$O($\nu, \nu^{\prime}$)$^{16}$O(12.97 MeV, $2^-$) with a water-Cherenkov detector in the supernova neutrino bursts}

%%%% To generate auto affiliation numbers please use \author{}\affil{} command

\author[1,*]{Makoto \textsc{SAKUDA}}
\affil[1]{Physics Department, Okayama University, Okayama 700-8530, Japan
\email{sakuda-m@okayama-u.ac.jp}}

\author[2,*]{Toshio \textsc{SUZUKI}}
\affil[2]{Department of Physics, College of Humanities and Sciences, Nihon University, Tokyo 156-8550, Japan
\email{suzuki.toshio@nihon-u.ac.jp}}

\author[3]{Mandeep Singh \textsc{REEN}}
\affil[3]{Department of Physics, Akal University, Punjab 151302, India}

\author[4]{Ken'ichiro \textsc{NAKAZATO}}
\affil[4]{Faculty of Arts and Science, Kyushu University, Fukuoka 819-0395, Japan}

\author[5]{Hideyuki \textsc{SUZUKI}}
\affil[5]{Department of Physics, Faculty of Science and Technology, Tokyo University of Science, Chiba 278-8510, Japan}

%\author[3]{Insert fourth author name here} %%% Use optional bracket [3] to change the respective address
%\affil{Insert third author address here}

%\author{Insert last author name here\thanks{These authors contributed equally to this work}}
%\affil{Insert last author address here}

%%% To include the collaborator name... Please use the command "\collaborator"
%%% For example: \collaborator{ATLAS Collaboration}

\begin{abstract}%
We first discuss and determine the isospin mixing of the two $2^-$ states (12.53 MeV and 12.97 MeV) of $^{16}$O nucleus using the inelastic electron scattering data. We then evaluate the cross section of 4.4-MeV $\gamma$ rays produced in the neutrino neutral-current (NC) reaction $^{16}$O($\nu, \nu^{\prime}$)$^{16}$O(12.97 MeV, $2^-$) in a water Cherenkov detector at the low energy below 100 MeV. The detection of $\gamma$ rays for $E_{\gamma}>5$~MeV from the NC reaction  $^{16}$O($\nu, \nu^{\prime}$)$^{16}$O($E_x>16$ MeV, $T=1$) with a water Cherenkov detector in the supernova neutrino bursts has been proposed and discussed by several authors previously. In this article, we discuss a new NC reaction channel from $^{16}$O(12.97 MeV, $2^-$) producing a 4.4-MeV $\gamma$ ray, the cross section of which is more robust and even larger at the low energy ($E_{\nu}<25$ MeV) than the NC cross section from $^{16}$O($E_x>16$ MeV, $T=1$). We also evaluate the number of such events  induced by neutrinos from supernova explosion which can be observed by the Super-Kamiokande, a 32 kton water Cherenkov detector in the Earth.
\end{abstract}

\subjectindex{C43, D02, D03, D21, F22}

\maketitle

\section{Introduction}

In the past, the neutral-current (NC) reactions in a few tens of MeV were observed in $^{12}$C($\nu, \nu^{\prime}$)$^{12}$C(15.1 MeV, $J^P=1^+$) by the KARMEN experiment~\cite{Karmen1,Karmen2} and recently in coherent $\nu$-CsI(Na) scattering by COHERENT Collaboration~\cite{Coherent}. Authors of Refs.~\cite{Kolbe, Kolbe1} proposed to measure $\gamma$ rays for the detection of NC inelastic events from $^{12}$C and $^{16}$O($\nu, \nu^{\prime}$) reactions induced by supernova (SN) neutrinos, and Beacom and Vogel~\cite{Beacom1, Kolbe02} estimated the number of NC $\gamma$-ray events with a water Cherenkov detector. 
%%Beacom, Farr, and Vogel~\cite{Beacom2} proposed to measure neutrino-proton elastic scattering ($\nu +p \rightarrow \nu +p$) in scintillator detectors~\cite{Dasgupta}. 
While the electron-neutrino $\nu_e$ and their antiparticle $\overline{\nu}_e$ interact through both charged-current (CC) and NC reactions in the SN core and in the neutrino detectors on the Earth, the muon-neutrino $\nu_{\mu}$ and the tau-neutrino $\nu_{\tau}$  (and their antiparticles) only interact through NC reactions, since their energies are too low to produce muons and tau-leptons through CC reactions. 
Since the neutrino spectrum from SN explosion was only measured for $\overline{\nu}_e$ at SN1987A and  the neutrino spectra for other neutrino flavors are not known,  it is important to estimate and measure as many NC reactions with good accuracy for the better understanding of core-collapse SN explosion and the neutrino oscillations~\cite{SNrev1,SNrev2,SNrev3,SNrev4}.

The Super-Kamiokande (SK) Collaboration summarizes the detection channels from SN neutrino bursts with a water Cherenkov detector in Ref.~\cite{SK-SN1, SK-SN2}. The inverse $\beta$ decay (IBD) reaction induced by $\bar{\nu}_e$ on a proton in water ($\bar{\nu_e}p\rightarrow e^+n$) is the dominant reaction. The detection of $\gamma$ rays for $E_{\gamma}>5$~MeV  from the NC reaction $^{16}$O($\nu, \nu^{\prime}$)$^{16}$O$(E_x>16\ {\rm MeV}, T=1$) with a water Cherenkov detector in the SN bursts was proposed by Langanke, Kolbe and Vogel~\cite{Kolbe1}  just when the SK experiment was about to start in 1996. The production of  the $\gamma$ rays for $E_{\gamma}>5$ MeV  from the NC reactions has been calculated and discussed previously since then by several authors~\cite{Kolbe1, Beacom1, Kolbe02}.  The neutrino-$^{16}$O charged-current (CC) cross section ~\cite{Haxton87, Kolbe02, Kolbe03} and its electron spectra~\cite{Nakazato1} in the SK experiment were studied for a SN neutrino burst in the previous publications.  

The SK experiment has been measuring the $^8$B solar neutrino spectrum since 1996~\cite{SKsolar1, SKsolar2, SKsolar3, SK-det}.  The analysis threshold of the recoil electron energy has recently become lower from 5 MeV (SK-I, II, III)~\cite{SKsolar1, SKsolar2, SKsolar3} to 3.5 MeV (SK-IV)~\cite{SK-det} by the  improvement of electronics system,  water system, calibration and analysis technique.  Thus,  it is timely to discuss a detection of the 4.4-MeV $\gamma$ rays. We  discuss for the first time a detection of 4.4-MeV $\gamma$ rays produced in the neutrino NC reaction $^{16}$O($\nu, \nu^{\prime}$)$^{16}$O$(12.97\ {\rm MeV}, 2^-$) with a water Cherenkov detector in the SN neutrino bursts.  The SK experiment,  now called SK-Gd~\cite{EGADS, SK-Gd1},  collects data with an addition of gadolinium to pure water  in order to observe the supernova relic neutrinos~\cite{Vagins, SK-SRN,KamLAND-SRN}.  

The following three quantities must be understood before the NC cross section of 4.4 MeV $\gamma$-ray production from $^{16}$O(12.97 MeV, $2^-$) is calculated: (a) the isospin mixing parameter between the 12.97 MeV and 12.53 MeV states from the ($e, e^{\prime}$) cross section, (b) the NC neutrino oxygen cross section  $^{16}$O($\nu, \nu^{\prime}$)$^{16}$O(12.97 MeV, $2^-$), and (c) the branching ratio Br$(\alpha_1)$ of the 12.97-MeV state decaying to $\alpha +^{12}$C(4.4 MeV, 2$^+$), which emits a 4.4-MeV $\gamma$ ray. Thus, this report is organized as follows. In Section~\ref{sec2}, we explain the isospin mixing  between the two $2^-$ states,  12.53 MeV  and 12.97 MeV. In Section~\ref{sec3}, we determine the isospin mixing parameter between the two states using the published ($e, e^{\prime}$) data.  In Section~\ref{sec4}, we show the NC cross section calculations.
% In Section V, we explain the typical neutrino flux spectra produced in core-collapse SN explosion.
%: one is used conventionally before and another is the one tabulated by Nakazato~${\it et\ al.}$~\cite{Nakazato}. 
In Section~\ref{sec5}, we combine the results of Sections~\ref{sec2}-\ref{sec4} with the typical neutrino flux spectra produced in core-collapse SN explosion and estimate the number of  $\gamma$-ray events produced in NC inelastic reactions from SN neutrinos.
% In Section VI, we combine the results of Sections II-V and estimate the number of  $\gamma$-ray events produced in NC inelastic reactions from SN neutrinos.
In addition, we summarize  in Appendix A all the formulas of the electron-oxygen inelastic reaction, the NC neutrino-oxygen cross section, the rate of muon capture on oxygen and $^{16}$N $\beta$-decay rate, which are used in this article.  In Appendix B, we determine the quenching factor $f_s=g_s^{\rm eff}/g_s$ of the spin g-factor $g_s$ for the magnetic form factors  and  the quenching factor $f_A=g_A^{\rm eff}/g_A$ of the weak axial-vector coupling  constant $g_A$ for the 12.97-MeV state, in order to evaluate both the electromagnetic and weak interactions with the 12.97-MeV state precisely.

\section{The two $2^-$ states, 12.53 MeV ($T=0$) and 12.97 MeV ($T=1$), of $^{16}$O and the isospin mixing between the two states}\label{sec2}

In this section, we explain the isospin mixing effect observed between the two $2^-$ states,  12.53 MeV ($T=0$) and 12.97 MeV ($T=1$), of $^{16}$O. 
The two excited states of $^{12}$C at 12.71 MeV ($1^+, T=0$) and 15.11 MeV ($1^+, T=1$) is a well-known example of the isospin mixing~\cite{Adelberger, Flanz, Cosel}. We start with a simple two-state model of the isospin mixing between the two adjacent states of a nucleus. The isospin mixing is known to be caused for example by the Coulomb interaction between the protons in the nucleus which may violate the isospin symmetry, but the exact origin of the effect is still unresolved~\cite{Auerbach1, Auerbach2}. The physical two $2^-$ states (the higher energy state $|U\rangle$ and the lower energy state $|D\rangle$) are written in terms of  the pure isospin states as,  
\begin{eqnarray}\label{isospin}
|U \rangle &=& \sqrt{1-\beta ^2} \,|U, T=1 \rangle -\beta \,|U,T=0 \rangle, \nonumber  \\ 
|D \rangle &=& \sqrt{1-\beta ^2} \,|D, T=0 \rangle +\beta \,|D,T=1 \rangle,
\label{eq:ud-states}
\end{eqnarray}  
where $|U\rangle$ and  $|D\rangle$ stand for the 12.97-MeV and 12.53-MeV states, respectively, and $\beta$ is the isospin mixing parameter. Another definition of the isospin mixing parameter $\epsilon$ is sometimes used and they are mutually related as $\beta=\epsilon /\sqrt{1+\epsilon ^2} $. 
 
The 12.97-MeV state and 12.53-MeV state lie just above the proton separation energy (12.1 MeV) of $^{16}$O. They both can decay to $p+^{15}$N(ground state (g.s.), $1/2^-, T=1/2$) and only the $T=0$ state can decay to $\alpha +^{12}$C(g.s., 0$^+$) and $^{12}$C(4.4 MeV, 2$^+$), whose energy threshold are 7.2 MeV and 11.6 MeV, respectively. If it were not for the isospin mixing effect, the 12.97-MeV state ($T=1$) would decay only to $p+^{15}$N(g.s.), which produces no singal in a water Cherenkov detector even if it is produced in the neutrino interactions.

There have been several reports on the isospin mixing between the 12.97 MeV ($T=1$) and 12.53 MeV ($T=0$) states previously~\cite{Stroetzel, Wagner, Leavitt, Zijderhand, Charity}.  Stroetzel~\cite{Stroetzel} measured the reduced transition probability $B(M2,q)$ of the two states in $^{16}$O($e, e^{\prime}$) reaction and discussed the effect of the isospin mixing between those two states, using the values $B(M2,q=0)$=0.42$\pm$0.10 fm$^4$ and 1.34$\pm$0.27 fm$^4$ for the 12.53 MeV and 12.97 MeV, respectively. He gave the ratio of $B(M2,q=0, 12.53\ {\rm MeV})/[B(M2,q=0, 12.53\ {\rm MeV})+B(M2,q=0, 12.97\ {\rm MeV})]$=0.25 as a size of the mixing effect. Wagner ${\it et\ al.}$~\cite{Wagner} measured the three  pickup reactions $^{17}$O($d$, $^3$He)$^{16}$N, $^{17}$O($d$, $^3 t$)$^{16}$O and $^{17}$O($^3$He, $\alpha$)$^{16}$O, and  they evaluated the isospin mixing between the 12.97 MeV state ($T=1$) and the 12.53 MeV state ($T=0$) to be $\epsilon ^2\ge$0.17$\pm$0.07, where the $\epsilon$ is the mixing parameter defined in the text after Eq.~(\ref{eq:ud-states}). We note that they need not only  the  $\alpha$ decay branching ratio  $\Gamma_{\alpha 1}/\Gamma$  but also the Coulomb penetration factor in order to determine  the isospin mixing parameter $\epsilon$ (or $\beta$)~\cite{Wagner, Adelberger, Leavitt}.  Wagner ${\it et\ al.}$ also quoted  the value of the mixing effect, 0.24$\pm$0.07, in their paper, by referring to the paper by Stroetzel~\cite{Stroetzel}. This value and the uncertainty, 0.24$\pm$0.07,  comes from the ratio 0.25 which is quoted by Stroetzel~\cite{Stroetzel}.
%, though the definition of the isospin mixing is not clear. %We note that the value can be calculated from $B(M2,q=0, 12.53 MeV)/[B(M2,q=0, 12.53 MeV)+B(M2,q=0, %12.97 MeV)]$, though the definition of the mixing is not clear.
   
Then, the two experiments measuring the proton capture reactions reported the branching ratio $\Gamma_{\alpha_1}/\Gamma={\rm Br}(12.97~{\rm MeV}\to \alpha + {}^{12}$C(4.4 MeV)): Leavitt~${\it et\ al.}$~\cite{Leavitt} gave $\Gamma_{\alpha _1}/\Gamma=0.37\pm0.06$, and Zijderhand and van der Leun~\cite{Zijderhand} reported $\Gamma_{\alpha_1}/\Gamma=0.22\pm0.04$; their measured values were very inconsistent with each other.  Recently, Charity ${\it et\ al.}$~\cite{Charity} measured the value $\Gamma_{\alpha_1}/\Gamma=0.46\pm0.08$ in the neutron-transfer reaction  using the ion beam, which is barely consistent with the former, but is inconsistent with the latter. The Evaluated Nuclear Structure File (ENSDF) evaluation \cite{ENSDF} and the compilation of Tilley ${\it et\ al.}$~\cite{Tilley} do not update the results yet. A problem of the variation in the measurements of $\Gamma_{\alpha_1}/\Gamma$ of $^{16}$O(12.97 MeV) remains unresolved until now. We do not discuss the variations of the three measurements which were observed by the hadronic reaction experiments~\cite{Leavitt, Zijderhand, Charity}, but we take a simple mean of the three branching ratio values and use this mean value, Br$(\alpha_1)=\Gamma_{\alpha_1}/\Gamma=0.35$, in the present paper\footnote{We note that such confusing situation is also seen in the isospin mixing of the two $J^P=1^+$  states (12.71 MeV and 15.11 MeV) of $^{12}$C. The analysis of Ref.~\cite{Cosel} shows that the variations of the isospin mixing effects between the two $1^+$ states (12.71 MeV and 15.11 MeV) of $^{12}$C  are seen in the hadronic reactions, but that the results are consistent with each other among ($e, e^{\prime}$) experiments~\cite{Flanz, Cosel}. The branching ratios $\Gamma_{\alpha_1}/\Gamma$ of 12.97 MeV and 12.53 MeV can be related to the isospin mixing parameter under the condition that the $\alpha$-particle penetrability from $^{16}$O at $E_x$=12.53 MeV and 12.97 MeV can be calculated precisely~\cite{Wagner, Leavitt}. }.
%In the present paper, we do not discuss the variations of the measured values which were observed by the hadronic reaction experiments~\cite{Wagner, Leavitt, Zijderhand, Charity}.
%\cite{Note2}.  
 
Since the NC $^{16}$O($\nu, \nu^{\prime}$)$^{16}$O(12.97 ${\rm MeV}, 2^-, T=1)$  cross section depends on the isospin mixing between the 12.97 MeV and 12.53 MeV, we evaluate the isospin mixing parameter using the existing $^{16}$O($e, e^{\prime}$) data of the two states in the next section.

\section{The evaluation of the isospin mixing parameter between the 12.53 MeV ($T=0$) and 12.97 MeV ($T=1$) of $^{16}$O}\label{sec3}
 
We re-examine the published ($e, e^{\prime}$) cross section at the 12.97-MeV and 12.53-MeV states~\cite{Stroetzel, Kim} and obtain the isospin mixing parameter $\beta$ defined in Eq.~(\ref{eq:ud-states}), using the squared root of the reduced transition probability $\sqrt{B(M2, q)}$.

We note that the 12.97-MeV state is the first strong $2^-$ excited state of $^{16}$O just above the proton separation energy and it is one of the dominant multipoles in the neutrino-oxygen interactions at low energy below 100 MeV. 
The electromagnetic form factors $F^2(q)$ of these states were measured in ($e,e^{\prime}$) reactions in 1960~\cite{Vanpraet,Stroetzel, Kim, Sick}. 
%%The reduced transition probabilities $B(M2,q)$ of those states were also measured by Stroetzel  and Kim ${\it et\ al.}$~\cite{Stroetzel, Kim}.  
Those data of the excited states from 12 MeV to 20 MeV of $^{16}$O were then examined by Donnelly and Walecka~\cite{Donnelly1, Donnelly2, Walecka75} and were applied to the calculations of the neutrino-oxygen cross section.  No new measurements of those states in ($e,e^{\prime}$) reactions have been performed since then. 
 
\begin{figure}[ht!]
\centering
\includegraphics[width=14.0cm,scale=1.0]{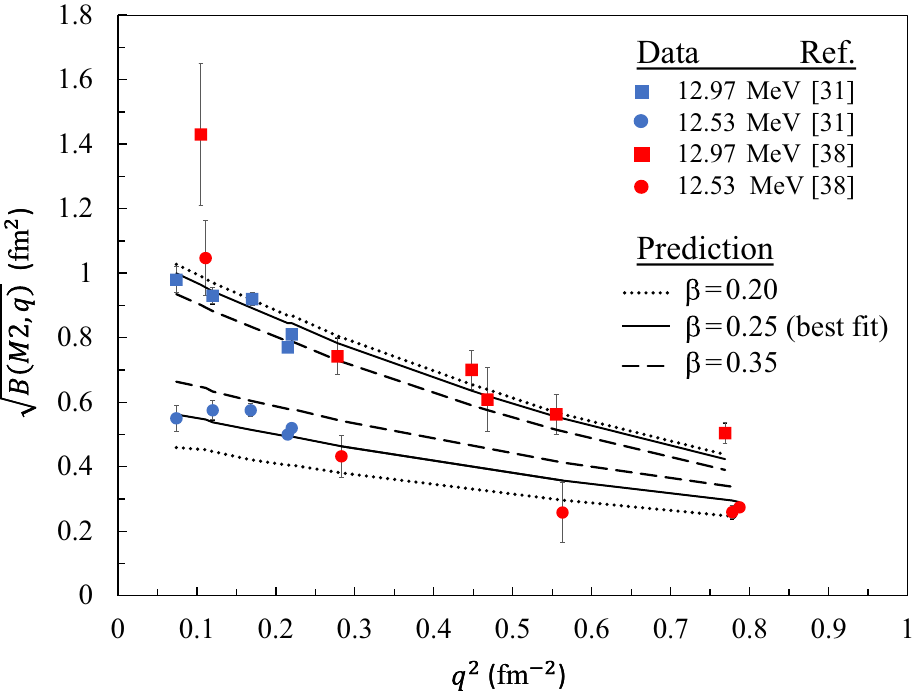}
\caption{The squared root of the reduced transition probability $\sqrt{B(M2,q)}$ for the excited states 12.97 MeV (closed squares) and 12.53 MeV (closed circles) as a function of the momentum transfer squared $q^2 ({\rm fm}^{-2})$. Data are taken from Stroetzel (blue data points, Ref.~\cite{Stroetzel}) and Kim ${\it et\ al.}$ (red data points, Ref.~\cite{Kim}). Predictions of $\sqrt{B(M2,q)}$ for 12.97-MeV and 12.53MeV states are shown for $\beta$=0.25 (best fit) as a upper solid line and a lower solid line, respectively. Predictions for $\beta$=0.2 (dotted lines) and 0.35 (dashed lines) are shown for comparison of the sensitivity of the $\beta$ values to $\sqrt{B(M2,q)}$. 
}\label{fig:redtp}
\end{figure}

In Appendix B, we evaluated the quenching factor $f_s$ of the spin $g$ factor and obtained $f_s=0.65$ at $E_x=13$~MeV region. We now use the data of the reduced transition probabilities $B(M2,q)$ for 12.53 MeV and 12.97 MeV using this $f_s$ value for various values of the isospin-mixing parameter $\beta$ defined in Eq.~(\ref{eq:ud-states}). 
Stroetzel published the data of $\sqrt{B(M2, q)}$~\cite{Stroetzel}.  We converted  the values of $F_T^2(M2, q)$ measured by Kim ${\it et \ al.}$~\cite{Kim} into $\sqrt{B(M2, q)}$ using Eq.~(\ref{eq:a12}) of Appendix A. The data of the squared root of the reduced transition probability $\sqrt{B(M2,q)}$ for the excited states 12.97 MeV (closed squares) and 12.53 MeV (closed circles) are plotted as a function of the momentum transfer squared $q^2 ({\rm fm}^{-2})$ are plotted in Fig.~\ref{fig:redtp}. Predictions of $\sqrt{B(M2,q)}$ for 12.97-MeV and 12.53MeV states are also shown for $\beta$=0.2 (dotted line), 0.25 (solid line) and 0.35 (dashed line) as a upper line and a lower line, respectively. We find that $\beta=0.25\pm0.05$ gives a good fit to the data.  
%Predictions for $\beta$=0.2 (dotted line) and 0.35 (dashed line) are shown for comparison of the sensitivity of the $\beta$ values to $\sqrt{B(M2,q)}$. 
%The $\sqrt{B(M2, q)}$ distributions are plotted in Fig.~\ref{fig:redtp} and we find that $\beta=+0.25\pm0.05$ gives a good fit to the data. 
The data of Sick ${\it et \ al.}$~\cite{Sick} for $q^2 >0.8$~fm$^{-2}$ are not used in this analysis, since their data contain significant contribution of 13.26 MeV ($3^-$) as shown in Fig.~\ref{FM2}. The data of Vanpraet~\cite{Vanpraet} are not used either, since the statistics is not good.

\section{Calculation of the cross section of the neutral-current neutrino oxygen reaction $^{16}$O($\nu, \nu^{\prime}$)$^{16}$O$(12.97~{\rm MeV}, 2^-$) and the 4.4-MeV $\gamma$-ray production}\label{sec4}

In the previous Sections~\ref{sec2} and \ref{sec3}, we fixed (a) the isospin mixing parameter $\beta=0.25$ between the 12.97 MeV and 12.53 MeV states and (c) the branching ratio ${\rm Br}(\alpha_1)={\rm Br}(12.97~{\rm MeV}\to \alpha +^{12}{\rm C}(4.4~{\rm MeV}))$=0.35. Since this paper concerns on the isospin mixing, good description of both $T=1$ and $T=0$ transition matrices of the electron-$^{16}$O and neutrino-$^{16}$O reactions is important. We thus use the shell-model calculation of Refs.~\cite{SFOtls, suzuki11} for the $T=1$ part of the Hamiltonian, called SFO-tls, and that of Ref.~\cite{YSOX} for the $T=0$ part of the Hamiltonian, called YSOX. 
The details of the shell model calculations are given elsewhere~\cite{SFO,suzuki06,TSuzuki2,YSOX} and all the formulas used in this paper are described in the Appendix A. 

\begin{figure}[ht!]
\centering
\includegraphics[width=12.0cm,scale=1.0]{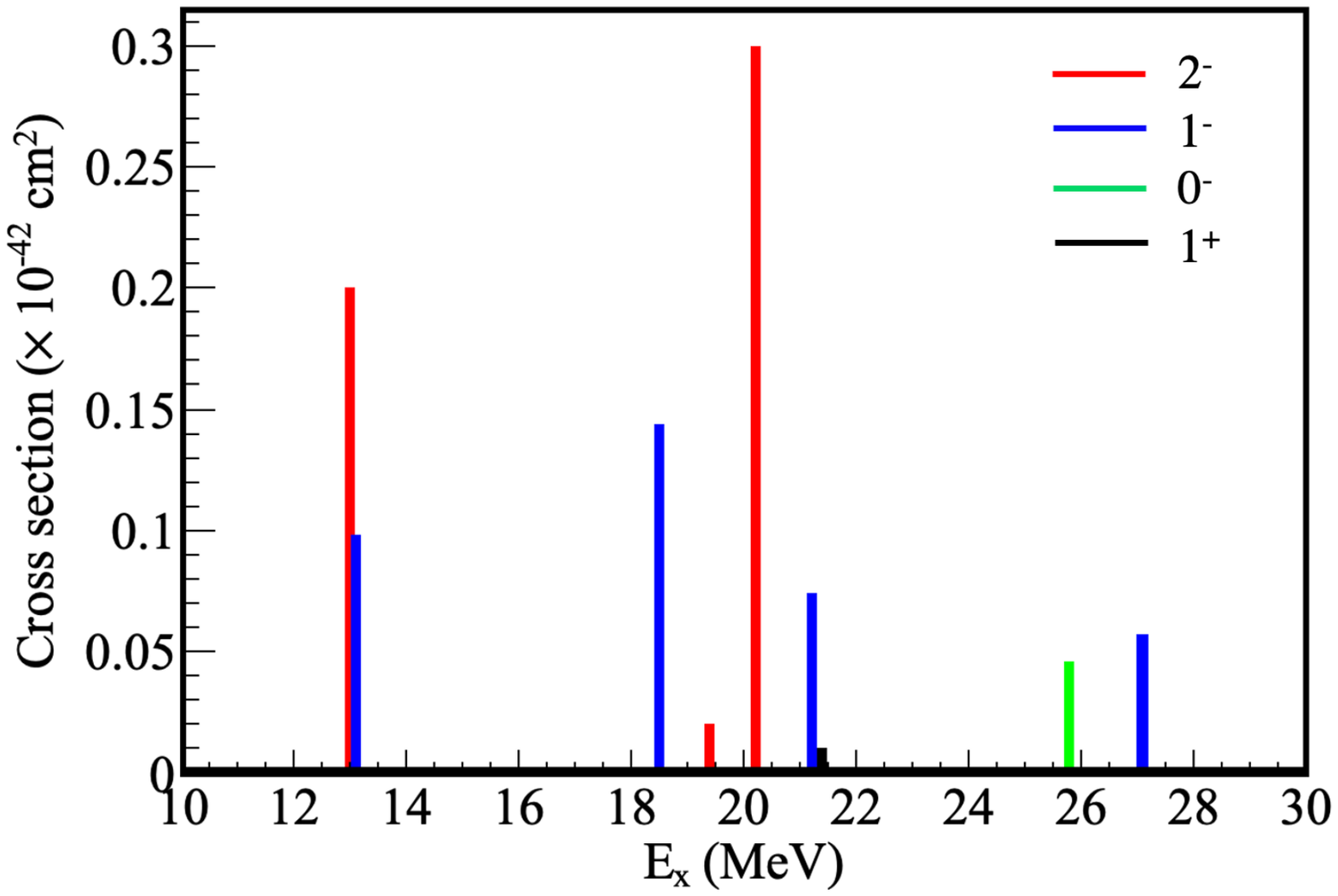}
\caption{The NC cross sections $\sigma (E_x)$ of  $\rm^{16}O(\nu,\nu^\prime) (10^{-42}\ cm^2)$ as a function of excitation energy $E_x$ at the neutrino energy  $E_\nu=30$ MeV. Multipoles of $J^P=2^-, 1^-, 0^-$ and $1^+$ are shown in red, blue, green and black bar graphs, respectively. The first peak (red bar) corresponds to the 12.97 MeV state ($2^-$, $T=1$). 
}\label{fig:dsdEx}
\end{figure}

Using the formula of Eq.~(\ref{eq:a2}), we calculate and show the NC cross section for the excited states of $^{16}$O at $E_{\nu}=30$~MeV in Fig.~\ref{fig:dsdEx}.  The multipoles ($J^P=1^-,\ 2^-$ and ${\rm 0^-}$, $T=1$)  via the spin-dipole transitions from the $p$-shell to the $sd$-shell are important. Since the transition strength of the multipole ($J$) is roughly proportional to $(2J +1)$, the contribution of the multipole (2$^-$) is the largest of the three multipoles. 
The first peaks are due to the excited states 12.97 MeV ($2^-$, $T=1$) and 13.09 MeV ($1^-, T=1$). The cross section of 12.79 MeV ($0^-$, $T=1$) is less than $0.01\times 10^{-42}\ {\rm cm}^2$ and is not seen. The 13.09 MeV state ($1^-$, $T=1$) decays only to the $T=1$ component of $p+^{15}$N(ground state (g.s.), $1/2^-$, $T=1/2$) due to the isospin conservation, giving no signal to a water Cherenkov detector.  The contribution of ${\rm 1^+}$ is smaller than that of ${\rm 0^-}$. The  NC cross sections calculated by the CRPA models~\cite{Kolbe, Jachowicz} show the larger contribution of the 13.09 MeV state ($1^-, T=1$) than that of the 12.97 MeV state ($2^-, T=1$). 

\begin{figure}[ht!]
\centering
\includegraphics[width=12.0cm,scale=1.0]{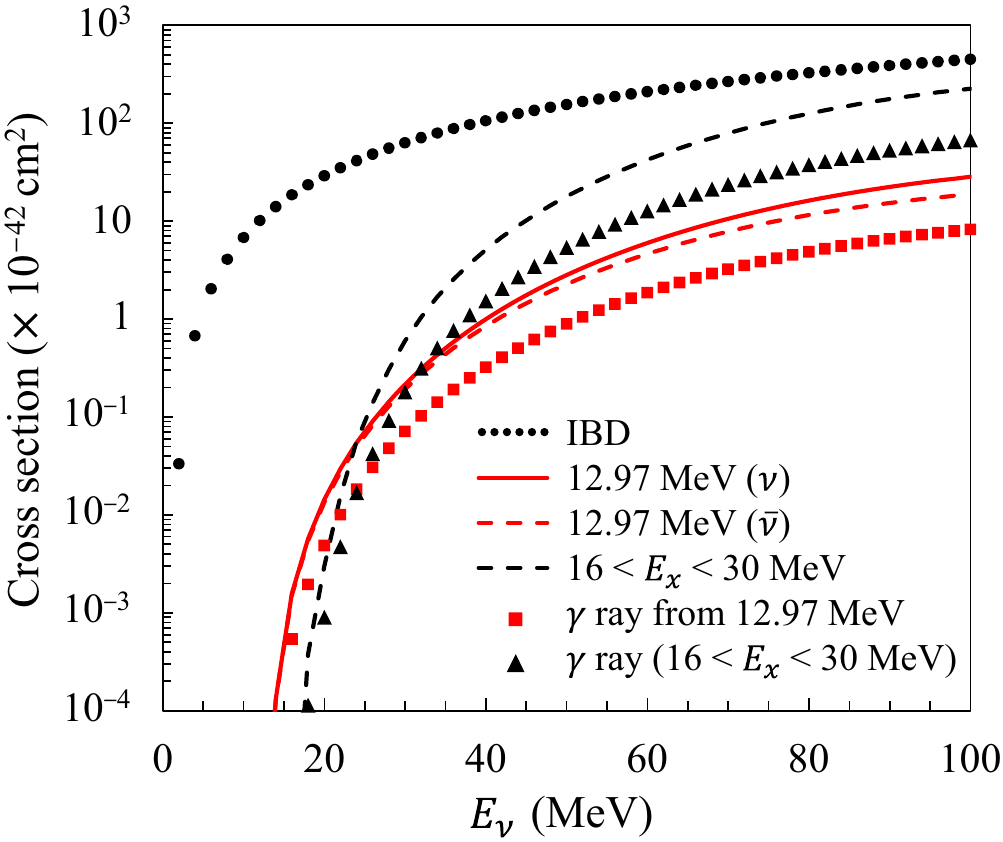}
\caption{The inelastic cross sections $^{16}$O($\nu,\nu^{\prime}$)$^{16}$O as a function of the neutrino energy  $E_\nu$.  
\label{fig:cross3}}
\end{figure}

The multipoles ($2^-, 1^-, 0^-, 1^+, T=1$) of the excitation energy $E_x$ above 16 MeV are also shown in Fig.~\ref{fig:cross3}. The shell model can predict the energies of the excited states of the multipoles ($2^-, 1^-, 0^-, T=1$) below 15 MeV within $\pm$1 MeV. But, since the excited states above 16 MeV overlap in data, the one to one correspondence between the observed states and the model predictions is difficult for $E_x>$ 16 MeV.

We show in Fig.~\ref{fig:cross3} the NC cross section of the 12.97-MeV state for a neutrino (red solid line) and an anti-neutrino (red dashed line) and the 4.4-MeV $\gamma$-ray production cross section (closed square symbol)  as a function of the neutrino energy. 
The neutrino cross section is larger by about 5\% at $E_{\nu}=20$ MeV and  by almost 50\% at $E_{\nu}=100$ MeV than the anti-neutrino cross section, since the interference term of Eq.~(\ref{eq:a2}) is proportional to $|q/(g_AM_N)|$ in the first approximation, where $q$, $g_A$ and $M_N$ are the momentum transfer, the weak axial-vector coupling constant and the mass of a nucleon, respectively~\cite{OConnell}. We show only the average cross section later for simplicity.
%Their calculation includes the cross section of the 12.97 MeV, the excited states above 16 MeV and the neutral-current quasi-elactic interaction. 
We also show the dominant IBD cross section~\cite{Strumia} (dotted line) and the NC cross section of the excited states  for $16<E_x<30$ MeV (including ${\rm 1^+}$) for comparison (dashed line). The calculation of $^{16}$O$(\nu,\nu^{\prime}X)$ cross section for $E_x>16$ MeV was shown to agree with that of by the CRPA model by Ref.~\cite{Kolbe03} within 10\% in the energy range $30<E_{\nu}<80$ MeV in Ref.~\cite{TSuzuki2}.
In order to compare the 4.4-MeV $\gamma$-ray production cross section (closed squares) with the latter cross section ($E_x >16$ MeV) producing $\gamma$ rays above 5 MeV (closed triangles), we multiply the latter cross section by the emission probability of about 30\% estimated by the authors of Ref.~\cite{Beacom1}\footnote{We did not use the simple analytic form $\sigma(E_{\nu})=(0.75\times 10^{-47}\ {\rm cm}^2)(E_{\nu}-15)^4$,  which is given in Ref.~\cite{Beacom1}, since it disagrees with the cross section ($\times$0.3) given by Ref.~\cite{Kolbe02}.}.
%, Note3}. 

First, we comment on the energy threshold and the energy dependence of the cross section of each reaction shown in Fig.~\ref{fig:cross3}. They are important features when we discuss about a detection of the SN neutrinos whose energy spectra may have a peak in the energy range between 10 and 15 MeV. The energy threshold of the IBD reaction is 1.8 MeV. The IBD cross section increases rapidly from 1.8 MeV to 10 MeV and moderately above 10 MeV.
The energy threshold of the NC cross section of the 12.97-MeV state is 12.97 MeV and if the neutrino energy is larger than 12.97 MeV, it can produce a 4.4-MeV $\gamma$ ray through its decay to $\alpha +^{12}$C(4.4 MeV) channel, the threshold of which is 11.6 MeV. The cross section increases rapidly from 12.97 MeV untill 100 MeV. The energy threshold of the NC cross section due to the excited states  ($E_x>16$ MeV) producing a 5.3-MeV or 6.3-MeV $\gamma$ ray is about 18 MeV. The cross section increases rapidly from 18 MeV untill 100 MeV.  

As shown in Figs.~\ref{fig:dsdEx} and \ref{fig:cross3}, the NC cross section of the 12.97-MeV state is larger than that of the excited states ($E_x>16$ MeV) for $E_{\nu} <25$ MeV. This feature is very important, since the majority of the neutrino spectrum from SN bursts is less than 25 MeV as shown later in Figs.~\ref{fig:KRJflux} and \ref{fig:NK1NK2flux}.  The latter cross section becomes larger than that of the 12.97 MeV state for $E_{\nu} >25$ MeV. 

The calculation of the NC cross section of the 12.97-MeV state is robust, since it is based on the measurements of both the electromagnetic form factors of 12.97 MeV and 12.53 MeV and the weak reaction rates (muon capture and $\beta$ decay), or, in other words, it is based on the quenching factors $f_s$ and $f_A$ for  the 12.97-MeV state and the isospin mixing parameter $\beta$ between 12.97-MeV and 12.53-MeV states as described in details in Appendix B and Section 3. However, the measurement of the branching ratio Br($\alpha_1$)=$\Gamma_{\alpha_1}/\Gamma$ of the 12.97 MeV state to produce 4.4 MeV $\gamma$ ray is uncertain, ranging from 0.22 to 0.46. 

The cross section of the NC reaction  $^{16}$O($\nu, \nu^{\prime}$)$^{16}$O($E_x>16$ MeV, $T=1$) can also produce $\gamma$ rays ($E_{\gamma}>5$ MeV),  but those excited states involve the several overlapping resonances and their form factors for the spin-dipole states ($1^-, 2^-$), which are important to neutrino interactions, are not clearly resolved yet~\cite{Kawabata} and the electromagnetic form factors not measured well in the electron scattering experiments~\cite{Kuchler, Wright}.  Furthermore, the $\gamma$-ray emission probabilities from those states ($E_x>16$ MeV) are not measured separately, except for a preliminary result for some of the unresolved states~\cite{Reen}.  As a result, the prediction of the cross section producing $\gamma$ rays from $E_x>16$ MeV is uncertain to $\pm$30\% or even more.  

 Finally, we comment on a small, but non-negligible, NC cross section $\sigma_{\rm NC}^D$ of the 12.53 MeV state (lower 2$^-$ state $D$) through of the isospin mixing, as compared to the NC cross section $\sigma_{\rm NC}^U$ of the 12.97 MeV (upper 2$^-$ state $U$). The ratio of $\sigma_{\rm NC}^D / \sigma_{\rm NC}^U$ is about  $\beta^2 /(1- \beta ^2)$=0.067  for $\beta$=0.25. However, the branching ratio of the 12.53 MeV state decaying to $\alpha$+$^{12}$C(4.4 MeV, 2$^+$)=0.83$\pm$0.03~\cite{ENSDF, Tilley}, which  is larger than the value 0.35 of the 12.97 MeV state decaying to $\alpha$+$^{12}$C(4.4 MeV, 2$^+$). The ratio of the 4.4-MeV production cross section from the two states is ($\sigma_{\rm NC}^D\times 0.83)/(\sigma_{\rm NC}^U \times 0.35)$=0.16. Thus, the 12.53 MeV state will add about 16\% to the NC cross section of the 4.4-MeV $\gamma$-ray production from the 12.97 MeV state at each $E_{\nu}$. We will not correct the numbers in the figures and tables for this effect in the present report, but we will report the details elsewhere~\footnote{M.Sakuda, T.Suzuki, M.S.Reen, K.Nakazato and H.Suzuki, manuscript in preparation.}.

%\section{Neutrino Flux from Supernova Explosion}
\section{Estimation of $\gamma$-ray production from neutral-current neutrino-oxygen inelastic reactions induced by supernova neutrinos}\label{sec5}

We evaluate the number of the NC events induced by neutrinos from SN explosion which can be observed by the Super-Kamiokande, a 32 kton water Cherenkov detector~\cite{SK-SN1, SK-SN2} in the Earth. 
While SN simulations with sophisticated neutrino interaction rates and multidimensional effects have been performed recently, differences in the time-integrated neutrino spectra are not drastic~\cite{OConnor, Just, Richers, Nagakura}. Thus, we  calculate the number of events at various average energies which we assume to be flavor independent. In this report, we adopt the following commonly used parametrization (called KRJ fit~\cite{Keil, Tamborra}) for the normalized neutrino spectra $f(E_\nu)$:
\begin{equation}
f(E_\nu) =  \frac{(\alpha+1)^{\alpha+1}}{\Gamma(\alpha+1) \langle E_\nu\rangle ^{\alpha+1}}E_\nu^\alpha\exp\Big{(}-\frac{(\alpha+1)E_\nu}{\langle E_\nu\rangle} \Big{)},
\label{eq:krj}
\end{equation}
where $\langle E_\nu\rangle$ is the average neutrino energy. In this expression, $\Gamma(\alpha+1)$ is the Gamma function and $\alpha$ is the pinching parameter. The distribution with $\alpha =2$ is referred to as Maxwell--Boltzmann (MB) thermal distribution and that with $\alpha =3$ is referred to as modified Maxwell--Boltzmann (mMB) distribution. As the value $\alpha $ becomes larger than 2, the high-energy tail of the distribution is more suppressed relative to the spectra of a thermal spectrum with the same average energy. The authors of Ref.~\cite{Keil} suggest the range $2<\alpha <4$ from the study of the neutrino spectra formation in a SN core, using Monte Carlo models.

\begin{figure}[ht!]
\includegraphics[width=14.0 cm,scale=1.0]{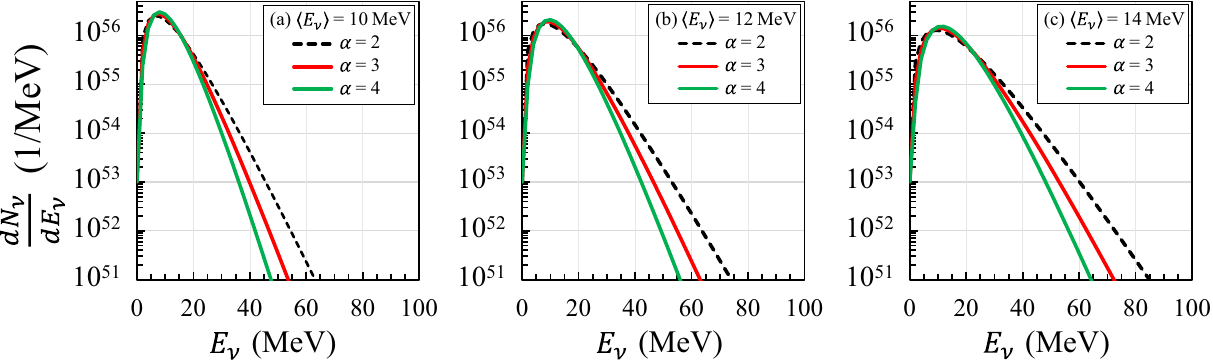}
\caption{The SN neutrino spectra for each neutrino flavor in the Keil-Raffelt-Janka (KRJ) parametrization with various $\langle E_\nu\rangle$ and $\alpha$. 
(a) $\langle E_\nu\rangle=10$ MeV and $\alpha=2$, 3 and 4, (b) $\langle E_\nu\rangle=12$ MeV and $\alpha=2$, 3 and 4, and (c) $\langle E_\nu\rangle=14$ MeV and $\alpha=2$, 3 and 4. \label{fig:KRJflux}} 
\end{figure}

%%\textcolor{green}{
The time-integrated number spectrum of neutrinos emitted from a SN core, $dN_{\nu}/dE_{\nu}$, is related to the normalized neutrino spectra $f(E_\nu)$ as
\begin{equation}
    \frac{dN_{\nu}}{dE_{\nu}} = \frac{E_\nu^{\rm tot}}{\langle E_\nu\rangle} f(E_\nu),
    \label{eq:tintspec}
\end{equation}
where $E_\nu^{\rm tot}$ is the total energy emitted by one neutrino flavor. Using the KRJ fit in Eq.~(\ref{eq:krj}), we plot $dN_{\nu}/dE_{\nu}$ in Fig.~\ref{fig:KRJflux} for various values of $\alpha$ and $\langle E_\nu\rangle$. Hereafter, we set $E_\nu^{\rm tot}=5\times 10^{52}$~erg for each neutrino flavor when we adopt the KRJ fit. The peak and the width of each spectrum is larger as the average value $\langle E_\nu\rangle$ takes a value from (a) 10 MeV to (c) 14 MeV. For the same value of $\langle E_\nu\rangle$, the high energy tail of the distribution for $E_{\nu}>25$~MeV is more suppressed as the pinching parameter $\alpha$ is larger. Furthermore, we calculate the number of events at various average energies using the corresponding neutrino flux $F(E_\nu)$ at a detector on the Earth, which is given as
\begin{equation}
    F(E_\nu) = \frac{1}{4\pi d_{\rm SN}^2}\frac{E_\nu^{\rm tot}}{\langle E_\nu\rangle} f(E_\nu).
    \label{eq:flux}
\end{equation}
Hereafter, we set the distance from a detector to the SN to $d_{\rm SN}=10$~kpc.

\begin{figure}[ht!]
\includegraphics[width=14.0 cm,scale=1.0]{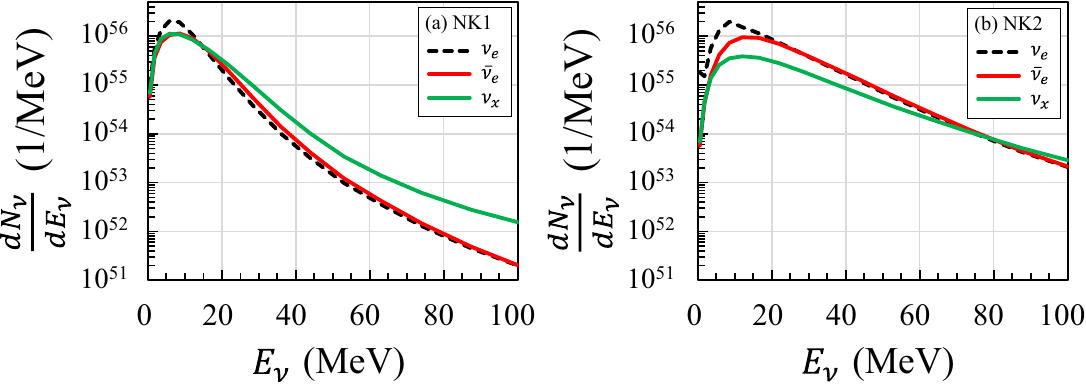}
\caption{(a) The SN neutrino spectrum from an ordinary SN model consistent with SN1987A (NK1) and (b) that from a black-hole-forming collapse (NK2), calculated for $\nu_e$, $\overline{\nu}_e$ and $\nu_x$ by Nakazato et al.~\cite{Nakazato}. 
 $\nu_x$ stands for either one of $\nu_{\mu}$, $\nu_{\tau}$ and their anti-particles. \label{fig:NK1NK2flux}}
\end{figure}

In addition to the KRJ fit with various $\alpha$ values, we choose other two sets  of the  SN neutrino flux spectra from the SN Neutrino Database~\cite{Nakazato}: the one is the model with  $(M,Z) = (20M_{\odot}, 0.02)$ and a shock revival time of 200 ms, which is chosen as an ordinary SN neutrino model consistent with SN1987A, and the another one  is the model with $(M,Z) = (30M_{\odot}, 0.004)$, which is a model of neutrino emission from a black-hole-forming collapse. Here, $M$, $M_{\odot}$ and $Z$ stand for a progenitor mass, a solar mass and the metalicity, respectively. We name the former spectra NK1 and the latter spectra NK2 in this report. In Figs.~\ref{fig:NK1NK2flux}(a) and (b), we show the time-integrated neutrino spectra $dN_{\nu}/dE_{\nu}$ of NK1 and NK2 models, respectively, for each neutrino flavor at a SN core. The average and total energies of the two models are listed in Table~\ref{tab:nkmodel}. The neutrino spectra of the NK1 and NK2 models have high energy tail in comparison with the models with the KRJ fit. These high energy components originate from the accretion phase, in which a SN core is heated due to the accretion of matter in the outer region. Since the average energy of neutrinos emitted from a SN core is time variant, the time-integrated neutrino spectra of the NK1 and NK2 models become less pinched. Whereas we assume the same average energy for each neutrino flavor for the models with the KRJ fit, the average energies of $\nu_x$ ($=\nu_\mu$, $\bar{\nu}_\mu$, $\nu_\tau$ and $\bar{\nu}_\tau$), $\bar{\nu}_e$, and $\nu_e$ are higher in that order for the models of NK1 and NK2. This is because, for $\nu_x$, the neutrino sphere resides deeper inside a SN core and the temperature at the neutrino sphere is higher. Note that, in this report, we do not take care of the effects of neutrino oscillation because the NC cross section is flavor independent in each of the neutrino and antineutrino sectors.

\begin{table}[ht]
\caption{Average energy, $\langle E_{\nu} \rangle$, and total energy, $E_{\nu}^{\rm tot}$, of the SN neutrino spectrum for $\nu_e$, $\bar{\nu}_e$, and $\nu_x$ ($=\nu_\mu$, $\bar{\nu}_\mu$, $\nu_\tau$ and $\bar{\nu}_\tau$).  
The neutrino spectra of the ordinary SN (NK1) and the case of a  black hole formation (NK2) are taken from Ref.~\cite{Nakazato}.}\label{tab:neutrino}
\centering
\begin{tabular}{ccccccc} 
 \hline
 \hline
& $\langle E_{\nu_e} \rangle$ & $\langle E_{\bar{\nu}_e} \rangle$ & $\langle E_{\nu_x} \rangle$ & 
$E_{\nu_e}^{\rm tot}$ & $E_{\bar{\nu}_e}^{\rm tot} $ & $E_{\nu_x}^{\rm tot}$  \\
neutrino flux model & (MeV) & (MeV)  & (MeV)  &  (10$^{52}$ erg)  &  (10$^{52}$ erg)  &  (10$^{52}$ erg)  \\
 \hline
Ordinary SN (NK1) & 9.32 & 11.1 & 11.9 &  3.30 & 2.82 & 3.27 \\
Black hole (NK2) & 17.5 & 21.7 &  23.4 & 9.49 &  8.10 & 4.00 \\
\hline
\hline
\end{tabular}
\label{tab:nkmodel}
\end{table}

We already showed in Fig.~\ref{fig:cross3} the NC cross section of the 12.97-MeV state and the 4.4-MeV $\gamma$-ray production cross section as a function of the neutrino energy and compared it with the sum of the cross section of the excited states (including ${\rm 1^+}$) for $16<E_x<30$~MeV. We denote this cross section as $\sigma _{\rm NC}^{(i)}(E_{\nu})$ for $E_x=12.97$~MeV ($i=1$) or for states in $16<E_x <30$~MeV ($i=2$). The corresponding $\gamma$-ray production cross sections are also denoted as $\sigma _{{\rm NC}\ \gamma}^{(i)}(E_{\nu})=\sigma _{\rm NC}^{(i)}(E_{\nu})\cdot R^{(i)}_{\gamma}$, where $R^{(i)}_{\gamma}$ denotes the $\gamma$-ray emission probability and we use $R^{(1)}_{\gamma}=0.35$ for $E_x=12.97$ MeV and $R^{(2)}_{\gamma}=0.30$ for $16<E_x <30$~MeV. 

We now calculate the number of the NC events containing 4.4-MeV $\gamma$ rays from $E_x=12.97$~MeV excited by neutrinos from SN explosion which can be observed by the Super-Kamiokande. We give the corresponding number of the NC events containing $\gamma$ rays from $16<E_x <30$~MeV for comparison. 
We fold the neutrino flux and the NC cross section to evaluate the number of NC $\gamma$ events as
\begin{eqnarray}
N_{{\rm NC}\ \gamma}^{(i)} = n_{\rm tar} \int_0^{E_\nu^{\rm max}} dE_\nu F(E_\nu) \sigma _{\rm NC\ \gamma}^{(i)}(E_{\nu}) ,
\label{eq:nc-gamma}
\end{eqnarray}
where $n_{\rm tar}$ is the number of targets ($\rm^{16}O$) in the neutrino detectors.
%$F(E_\nu) $ is the neutrino flux which is generated by SN and observed by a detector, $\sigma _{{\rm NC}\ \gamma}^{(i)}(E_{\nu})$ ($i=1,2$) is the NC $\gamma$-ray production cross section from the 12.97 MeV state and $E_x>$16 MeV at the incident neutrino energy $E_{\nu}$, respectively.
 
\begin{figure}[h!]
\centering
\includegraphics[width=14.0 cm,scale=1.0]{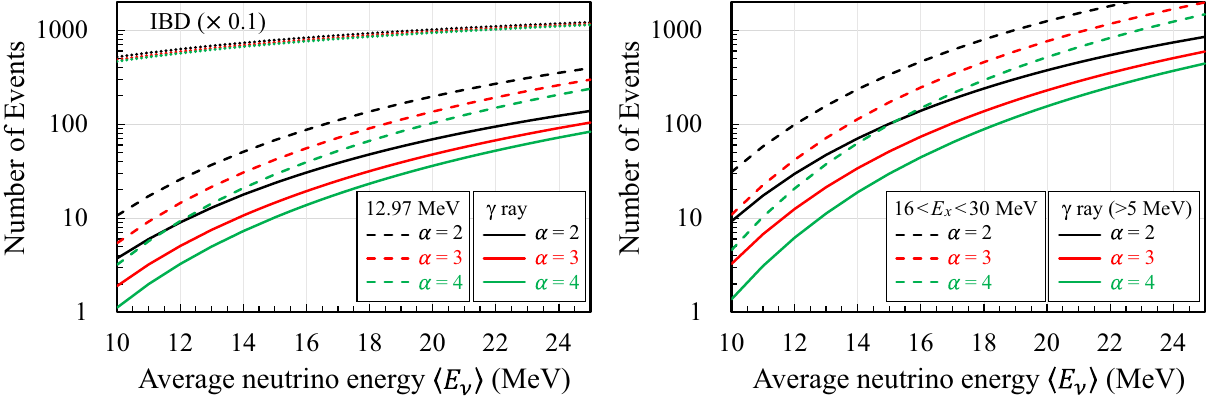}
\caption{Left: the number of events from $\rm^{16}O(\nu,\nu^\prime)^{16}O$(12.97 MeV) (dashed lines) and that producing 4.4 MeV $\gamma$ rays (solid lines)
as a function of the average neutrino energy, $\langle E_\nu\rangle$. The three (solid and dashed) lines correspond to the KRJ parameters with $\alpha=2$ (black line), 3 (red line), and 4 (green line), respectively. The corresponding numbers of the IBD events ($\times$0.1) are also shown in the figure, where the three lines almost overlap. Right: the corresponding number of events from  $\rm^{16}O(\nu,\nu^\prime)^{16}O$($16<E_x<30$ MeV) (dashed lines) and that producing $\gamma$ rays above 5 MeV (solid lines) are shown.  
\label{fig:MBmMB}}
\end{figure}

The results of Eq.~(\ref{eq:nc-gamma}) are shown in Fig.~\ref{fig:MBmMB} with the flux of the KRJ fit as a function of $\langle E_\nu \rangle$ for various $\alpha$ values. As expected, the number of NC $\gamma$ events is larger for higher $\langle E_\nu \rangle$. We can see that the number of NC $\gamma$ events is sensitive to the $\alpha$ value. Provided that the average neutrino energy $\langle E_\nu\rangle$ is the same, the spectrum is broader and neutrinos with $E_{\nu}>25$~MeV are more abundant for smaller $\alpha$ (see Fig.~\ref{fig:KRJflux}). Thus, the number of events is larger for smaller $\alpha$. This trend is common to the production of both the 4.4-MeV $\gamma$ rays and the $\gamma$ rays above 5 MeV. In contrast, the number of the IBD events, which is also shown in comparison with NC $\gamma$ events in Fig.~\ref{fig:MBmMB}, is rather insensitive to the $\alpha$ values.  We again note that the NC neutrino-$^{16}$O inelastic cross section increases exponentially as $E_{\nu}$ exceeds the threshold energy $E_x=12.97$ MeV and excited states $E_x>16$ MeV as shown in Fig.~\ref{fig:cross3}. The cross section of IBD reaction increases rapidly above the threshold energy (1.8 MeV) and moderately above 20 MeV.

\begin{figure}[h!]
\centering
\includegraphics[width=12.0cm,scale=1.0]{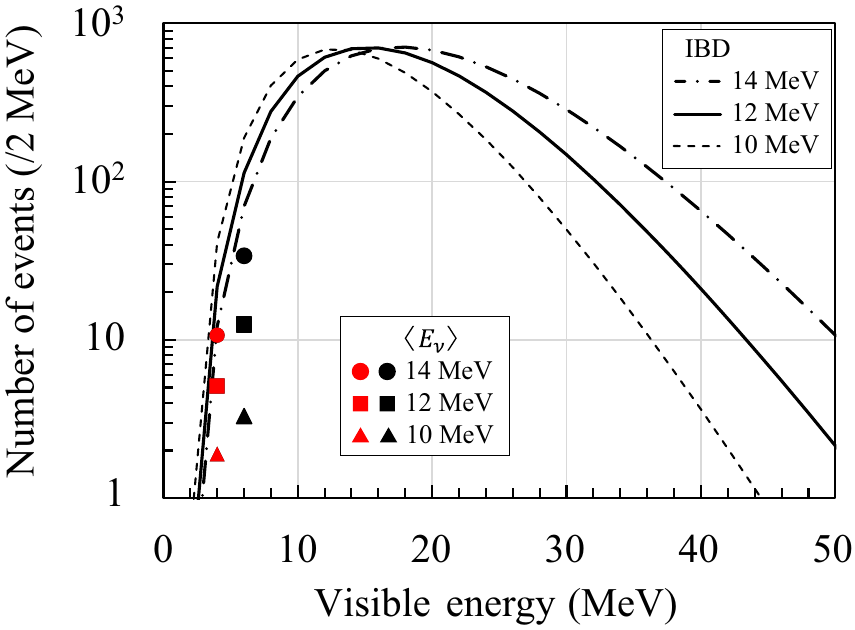}
\caption{Visible energy spectrum of the $\gamma$-ray (4.4 MeV from 12.97 MeV, red closed symbols, and 6.3 MeV (and 5.3 MeV) from $16<E_x<30$ MeV, black closed symbols) and the IBD events (black lines), expected at SK detector. The number of events (/2 MeV) corresponds to the three cases with $\alpha=3$ and $\langle E_\nu\rangle=10$ MeV, 12 MeV and 14 MeV. 
\label{fig:SKspectra}}
\end{figure}

In Fig.~\ref{fig:SKspectra}, the number of NC $\gamma$ events is compared with the IBD event spectrum  with an energy-bin width of 2 MeV as a function of the visible energy $E_{\rm vis}$, where the KRJ fit is adopted with $\alpha=3$ and $\langle E_\nu\rangle=10$, 12 and 14 MeV. Incidentally, the values of the corresponding event numbers are shown in Table~\ref{tab:evnmb1}. We note again that the numbers of NC $\gamma$ events are larger as the $\alpha$ value becomes smaller, or equivalently the higher energy tail of the SN spectra increases.  %%\textcolor{red}{20221005 Sakuda added: 
We assume that the IBD events can be identified and reconstructed unambiguously by the delayed coincidence method in the SK-Gd detector and that the $\bar{\nu}_e$ spectrum can be measured from the visible energy using the relation $E_{\bar{\nu}_e}=E_{\rm vis}+1.8$ MeV as described in Ref.~\cite{SK-SRN}. 
%% Sakuda comment out since IBDs are measured: In such a low-energy regime (around 4.4 MeV), the contribution of NC $\gamma$ events may be comparable with the IBD events.
On the other hand, the NC $\gamma$ events from the 12.97-MeV state  produce the visible energy due to a 4.4 MeV $\gamma$ ray, regardless of the incident neutrino energy. We also assume that the $\gamma$-ray energy can be reconstructed as the visible energy $E_{\rm vis}$. We plot the neutrino energy $E_{\bar{\nu}_e}$ for the IBD events in Fig.~\ref{fig:SKspectra}. We also show the $\gamma$-ray energy originating from NC $\gamma$ events of the excited states $^{16}$O($16<E_x<30$ MeV), assuming that the detector can measure the total energy of all $\gamma$ rays even in a cascade transition and the number of events is plotted at the visible energy at 6.3 MeV, representing 5.3 MeV, 6.3 MeV and other $\gamma$ rays.

\begin{table}[h]
\centering
\caption{Expected number of neutrino events from a core-collapse SN at 10 kpc to be detected at Super-K (32~kton) for the models with KRJ fit.}
\begin{tabular}{cccc} 
\hline
 \hline
KRJ parameters ($\alpha,\  \langle E_{\nu} \rangle$ [MeV])  &  (3, 10)  &  (3, 12) & (3, 14)  \\
\hline
 $p(\bar{\nu}_e,e^+)n$ & 4840  & 5900 & 6900 \\
 \hline
% NC $\rm^{16}O(\nu,\nu^\prime)\rm^{16}O^*(12.97MeV)$, $E_{\gamma}=$4.4 MeV) &  1.9 & 5.1 & 10.7 & 13 & 158 \\
$^{16}$O($\nu$, $\nu^{\prime}$)$^{16}$O(12.97 MeV), $E_{\gamma}=4.4$ MeV &  1.9 & 5.1 & 10.7 \\
$^{16}$O($\nu$, $\nu^{\prime}$)$^{16}$O($16<E_x<30$ MeV), $E_{\gamma}>5$ MeV~\cite{TSuzuki2,Kolbe02} &  3.3 & 12.5 & 33.9 \\
 \hline
 \hline
\end{tabular}
\label{tab:evnmb1}
\end{table}

Besides the IBD events, the elastic neutrino-electron scattering and the CC reactions on $^{16}$O produce possible background events for the NC $\gamma$ events. The numbers of those events were calculated for NK1 and NK2 models in Ref.~\cite{Nakazato1}. While they depend on the neutrino mass hierarchy through the neutrino oscillation, the dependence is insignificant and one can refer to Ref.~\cite{Nakazato1} for the details. In Table~\ref{tab:evnmb2}, they are compared to the numbers of the NC $\gamma$ events calculated in the present work. Whereas the CC reactions provide the larger number of events than the NC reactions, the visible energy of those events due to a scattered electron/positron will spread from 0 to 100 MeV and the number of events around 4 to 6 MeV in the visible energy is much smaller than that of the NC $\gamma$ events (see Figs.~5 and 6 of Ref.~\cite{Nakazato1}). We can also see that the NK2 model, which is for the black-hole-forming collapse, has the much larger number of NC $\gamma$ events than the NK1 model, which is for the ordinary core collapse SN. This is because the neutrino spectrum of the NK2 model has not only higher average and larger total energies (Table~\ref{tab:nkmodel}) but also the longer high-energy tail (Fig.~\ref{fig:NK1NK2flux}) than the NK1 model. Note that, in the case of the black hole formation, the mass accretion continues longer and more high energy neutrinos are emitted in comparison with the NK1 model.

\begin{table}[h]
\centering
\caption{Expected number of neutrino events from a core-collapse SN at 10 kpc to be detected at Super-K (32~kton) for the NK1 and NK2 models.}
\scalebox{0.85}{
\begin{tabular}{ccccccc} 
\hline
 \hline
Model  &  NK1  &  NK1 & NK1 & NK2 & NK2 & NK2  \\
Neutrino oscillation  &  No osc. &  NH & IH & No osc. & NH & IH \\
\hline
Previous work~\cite{Nakazato1} & & & & & \\
 $p(\bar{\nu}_e,e^+)n$ & 3199  & 3534 & 4242 & 17525 & 14879 & 9255 \\
  $\nu$e elastic scattering  & 140 & 157 & 156 & 514 & 320 & 351  \\
  $^{16}$O($\nu_e$, $e^-$)+$^{16}$O($\bar{\nu}_e$, $e^+$), $E_e>5$ MeV~\cite{TSuzuki2} & 77 & 236 & 237 & 3831 & 3607 & 3448 \\
\hline
% NC $\rm^{16}O(\nu,\nu^\prime)\rm^{16}O^*(12.97MeV)$, $E_{\gamma}=$4.4 MeV) &  1.9 & 5.1 & 10.7 & 13 & 158 \\
Present work & & & & & \\
$^{16}$O($\nu$,$\nu^{\prime}$)$^{16}$O(12.97 MeV), $E_{\gamma}=4.4$ MeV & 20 & 20 & 20 & 240 & 240 & 240 \\
$^{16}$O($\nu$,$\nu^{\prime}$)$^{16}$O($16<E_x<30$ MeV), $E_{\gamma}>5$ MeV~\cite{TSuzuki2,Kolbe02} & 140& 140& 140& 984& 984& 984 \\
 \hline
 \hline
\end{tabular}
}
\label{tab:evnmb2}
\end{table}

\begin{figure}[h!]
\centering
\includegraphics[width=10.0cm,scale=1.0]{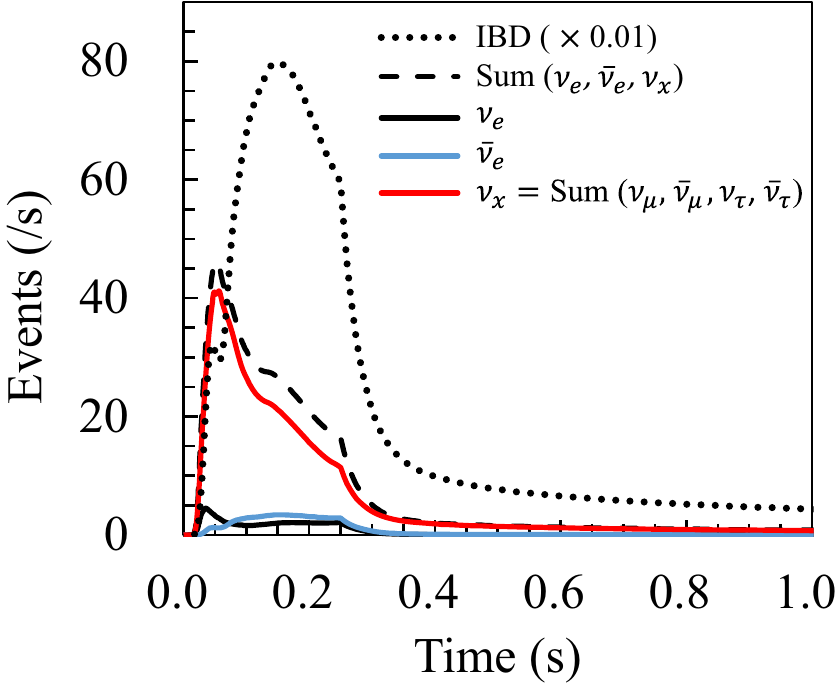}
\caption{The number of events per second from $\rm^{16}O(\nu,\nu^\prime)^{16}$O(12.97 MeV) producing 4.4 MeV $\gamma$ rays as a function of time for the NK1 model. The time is measured from the bounce of a SN core. The black dashed line denotes the contribution from all neutrinos. The black and blue lines denote the contribution from neutrinos emitted as $\nu_e$ and $\bar{\nu}_e$ at the source, respectively. The red line denotes the contribution from neutrinos emitted as $\nu_{\mu}$, $\bar{\nu}_{\mu}$,  $\nu_{\tau}$, and $\bar{\nu}_{\tau}$. The black dotted line denotes the IBD events ($\times$0.01), where the effects of the neutrino oscillation are not considered.
\label{fig:EventsvsTime}}
\end{figure}

In Fig.~\ref{fig:EventsvsTime}, the number of events per second in the NK1 model is shown for NC events with 4.4~MeV $\gamma$ rays. For comparison, the number of the IBD events calculated using the cross section from Ref.~\cite{Strumia} is also shown, while the effect of the neutrino oscillation was not considered. We can see that the rate of NC $\gamma$ events is higher for the accretion phase ($\lesssim$0.3~s in Fig.~\ref{fig:EventsvsTime}) than for the cooling phase ($\gtrsim$0.3~s). Furthermore, the contribution of neutrinos that were emitted as $\nu_x$ at the source is large because their average energy is high as already stated (Table~\ref{tab:nkmodel}). Therefore, the detection of NC $\gamma$ events is advantageous to probing $\nu_x$ emission in a SN core \cite{Kolbe1,Beacom1}. In addition, since the flux spectra and time variation of emitted neutrinos are different among models \cite{Nakazato,Jost}, the detection of NC $\gamma$ events would be useful for the model discrimination.

The detection of a single 4.4 MeV $\gamma$ ray in a water Cherenkov detector is a challenge, but the recent Super-Kamiokande detector  reports an analysis of the recoil electron kinetic energy  above 3.5 MeV threshold~\cite{SK-det}. Further, since one knows the start time of the SN neutrino bursts by the dominant IBD signals, one may be able to detect this 4.4-MeV signals using this start time, just as the T2K collaboration reported the detection of  6-MeV $\gamma$ rays from the NC neutrino-oxygen quasielastic interaction~\cite{Artur} with the Super-Kamiokande detector by knowing the GPS start time of the signal arrival~\cite{T2K6MeV1,T2K6MeV2}.

%%%%%%%%%%%%%%%%%%%%%%%%%%%%%%%%%%%%%%%%%%%%%\fi

\section{Summary}\label{sec6}
%\textcolor{red}{Sakuda, 2022.1031: 
We have re-examined the published electromagnetic form factors of 12.97 MeV and 12.53 MeV states of $^{16}$O considering the isospin mixing effect and determined both the quenching factor for the  spin $g$ factor and the isospin-mixing parameter $\beta$ of the two $2^-$ states to be $f_s=0.65\pm 0.05$ and $\beta=+0.25\pm 0.05$, respectively. We then determined the quenching factor for the axial-vector coupling constant  to be $f_A=0.68\pm 0.05$.  It is of fundamental importance to nuclear physics to understand the isospin mixing in light nuclei.
We note that the first precise determination of the electroweak matrix elements of the 13 MeV complex was performed by Donnelly and Walecka~\cite{Donnelly1,Donnelly2,Walecka75}, without the effect of isospin mixing.

Using the above values, we discussed a new NC reaction channel from $^{16}$O(12.97 MeV, $2^-$), the cross section of which is more robust and even larger at the low energy ($E_{\nu}<25$ MeV) than the NC cross section from O$(E_x>16~{\rm MeV}, T=1$). The accuracy of  our calculation for the new channel is about 20\%. The threshold neutrino energy of this new reaction is as low as 12.97 MeV, which is about the peak neutrino energy of the neutrino spectrum expected for the supernova explosion. The threshold energy of the  NC $\gamma$ events from excited states ($E_x>$ 16 MeV) is about 18 MeV. The cross section increases rapidly above the threshold energy as the neutrino energy increases.

We point out a possible detection of 4.4-MeV $\gamma$ rays produced in the neutrino NC reaction $^{16}$O($\nu, \nu^{\prime}$)$^{16}$O(12.97 MeV, $J^P=2^-$, $T=1$) with a water Cherenkov detector in the SN neutrino bursts. This reaction from the 12.97-MeV state produces the visible energy of about 4.4 MeV in the detector, regardless of the incident neutrino energy, while the IBD events and the CC neutrino-oxygen reaction produce the visible energy of the electron and positron related to the incident neutrino energy. The 4.4-MeV $\gamma$-ray emission probability Br($\alpha_1)$ of $^{16}$O(12.97 MeV, $J^P=2^-$) is still uncertain to 50\%, since there are three conflicting measurements which were performed by the hadronic reaction experiments~\cite{Leavitt, Zijderhand, Charity}.

We also evaluated the expected number of NC $\gamma$ events in the Super-Kamiokande detector~\cite{SK-SN1,SK-SN2} due to the new NC reaction channel for typical models of supernova neutrinos. One is the flux model called the Keil-Raffelt-Janka (KRJ) fit~\cite{Keil, Tamborra} with various average energies $\langle E_\nu\rangle$ and various pinching parameters $\alpha$, where we assume the flux to be flavor independent. We found that the number of NC $\gamma$ events from this channel is larger for higher $\langle E_\nu \rangle$. We showed, as an example, that the number of NC $\gamma$ events from this channel changes from 1.9 to 10.9 events when $\langle E_\nu\rangle$ changes from 10 MeV to 14 MeV for a fixed value $\alpha$=3, while the number of IBD events changes from 4840 to 6900 events. As for dependence on $\alpha$ values, we found that the smaller $\alpha$ is, the larger the number of NC $\gamma$ events is. It is because neutrinos with $E_{\nu}>25$~MeV are more abundant for smaller $\alpha$ models with given $\langle E_\nu \rangle$ and $E_{\nu}^{\rm tot}$.

Another set of the  SN neutrino flux spectra (NK1 and NK2) are taken from the SN Neutrino Database~\cite{Nakazato}. NK1 is chosen as an ordinary SN neutrino model consistent with SN1987A and NK2 is a model for a black-hole-forming collapse. We calculated the numbers of NC $\gamma$ events from the new reaction channel in Table~\ref{tab:evnmb2} for NK1 and NK2 models~\cite{Nakazato1}. Since the NC reaction is independent of the neutrino oscillations, the numbers of expected events are the same for the non-oscillation case and the oscillation cases (NH and IH). We included the number of the CC events in the Table for comparison. The number of NC $\gamma$ events for NK2 (black-hole-forming case) is  much larger than that for NK1, since the neutrino spectrum of the NK2 model has not only larger total energies but also  larger high-energy tails resulting in higher average energies. 
We also showed the number of NC $\gamma$ events from the new reaction channel per second in the NK1 model and compared it with that of the CC IBD events, while the effect of the neutrino oscillation was not considered. We find that the rate of NC $\gamma$ events is higher for the accretion phase than for the cooling phase. Furthermore, the contribution of $\nu_x$ neutrinos at the source is large because their average energy is high within this model. It suggests that the detection of the new channel may be useful to probing $\nu_x$ emission in a SN core and also useful for the model discrimination.

We hope that new measurements of the cross section of  $^{16}$O($e,e^{\prime}$)$^{16}$O(12.53 MeV, 12.97 MeV, $J^P=2^-$) and the branching ratios of $^{16}$O(12.53 MeV, 12.97 MeV, 2$^-$) decaying to $p$ and $\alpha$ channels  will be performed in the near future at the low-energy  electron accelerators ($E_e$=30-80 MeV) at S-DALINAC~\cite{Cosel2016}, at MESA accelerator~\cite{MESA},  or at the Research Center for Electron-Photon Science (Tohoku University)~\cite{Suda1, Suda2}, so that the isospin mixing of the two $2^-$ states can be measured accurately and  the prediction of the NC neutrino-oxygen cross sections for the 12.97 MeV state and the subsequent 4.4-MeV $\gamma$-ray production can be accurate to a level of 10\% or less.

\section*{Acknowledgment}
One of the authors (T.S.) would like to thank Prof. K. Langanke for a valuable comment. This work was partially supported by JSPS Grant-in-Aid for Scientific Research (No. 19K03855 and No. 20K03989) and also by JSPS Grant-in-Aid for Scientific Research on Innovative areas "Unraveling the History of the Universe and Matter Evolution with Underground Physics" (No. 19H05802 and No. 19H05811).

\appendix{
\section*{Appendices}

\section{Shell model calculations of $^{16}$O($e,e^{\prime}$) cross section, $^{16}$O($\nu,\nu^{\prime}$) cross section, the muon capture on $^{16}$O and the $\beta^-$-decay from $^{16}$N to $^{16}$O}

The calculations of the cross section of the electron-$^{16}$O scattering, the rate of the muon  capture on $^{16}$O and the $\beta^-$ decay from the ground state ($2^-$, $T$=1) of $^{16}$N to the ground state ($0^+$)  of $^{16}$O, and  the cross section of the NC neutrino-$^{16}$O reaction are based on the shell-model Hamiltonian, called SFO-tls~\cite{SFOtls, suzuki11}. This shell model improved the $p$-$sd$ cross shell part of the shell model, called SFO shell model~\cite{SFO}, which was shown to reproduce the neutrino-$^{12}$C charged-current reaction cross sections measured by the KARMEN and LSND experiments~\cite{KarmenCC1, KarmenCC2, LSNDCCe,LSNDCCm1,LSNDCCm2}. Thus, this SFO-tls model works better for the spin-dipole interactions between the $p$-$sd$ cross shell which is important in neutrino-$^{16}$O reactions, than the SFO model, while keeping the good feature of the Gamov-Teller interactions.

Though the $T$=1 states of $^{16}$O are well described by the SFO-tls shell model, the $T$=0 states of $^{16}$O, such as the energy levels, are shown to be better described by another shell model, called YSOX model~\cite{YSOX}.  We thus use the shell model calculation of SFO-tls for the $T$=1 part of the Hamiltonian and that of YSOX for the $T$=0 part of the Hamiltonian. 
%%\textcolor{red}{Sakuda, 2022.0604: I do not know much of YSOX. I would like to ask Suzuki-san to add a few lines of texts concerning YSOX and a consistency with other models (WBP, WBT). }
%\textcolor{red}{Suzuki, 2022.0707:
The YSOX Hamiltonian~\cite{YSOX} constructed in $p$-$sd$ shell developed from a monopole-based universal interaction ($V_{\rm MU}$)~\cite{VMU}. 
Its $p$-$sd$ cross shell part has the same tensor and two-body spin-orbit components of meson-exchanges as the SFO-tls, while the central components have been tuned based on those of $V_{\rm MU}$. The YSOX can reproduce well the ground-state energies, energy levels, electric quadrupole properties, and spin properties of boron, carbon, nitrogen and oxygen isotopes.
The monopole terms of the YSOX and SFO-tls are rather close to each other in the $T$=1 channel, but there are noticeable differences between them in the $T$=0 channel. 
The monopoles terms of the YSOX in the $T$=0 channel are generally close to those of the WBP and WBT~\cite{WB1992}, which are phenomenologically good Hamiltonians. 
Thus, the spectroscopic properties in the $T$=0 channel are expected to be generally favored for the YSOX compared with the SFO-tls. 
All the formulas used in this paper are described here.  
Since we focus on the isospin mixing between the two states (12.53 MeV and 12.97 MeV) with isospin $T$=0 and 1, we show the isospin structure of the hadronic currents explicitly, which consist of the electromagnetic current ($J_{\mu}^{(\gamma)}$), the neutrino charged current ($J_{\mu}^{\rm CC}$) and the neutral current ($J_{\mu}^{\rm NC}$). They are written as, 
\begin{eqnarray}
J_{\mu}^{\gamma} &=& J_{\mu}^{\gamma (V)}+J_{\mu}^{\gamma (S)},\  \nonumber \\
J_{\mu}^{\rm CC} &=& V_{\mu}^{\pm 1}+A_{\mu}^{\pm 1},\  \nonumber \\
J_{\mu}^{\rm NC} &=& V_{\mu}^{3}+A_{\mu}^{3}+A_{\mu}^{S}-2\sin^2\theta _W J_{\mu}^{\gamma}, 
\label{eq:a1}
\end{eqnarray}
where V and S denote the iso-vector ($T$=1) and iso-scalar ($T$=0) component, respectively, $V_{\mu}^i$ and $A_{\mu}^i$ ($i$=1-3) denote an iso-vector component of the weak vector and weak axial-vector current, respectively, $V_{\mu}^{\pm 1}=V_{\mu}^1\pm iV_{\mu}^2$  and $A_{\mu}^{\pm 1}=A_{\mu}^1\pm iA_{\mu}^2$. $A_{\mu}^S$ denotes an iso-scalar component of the weak axial current.  $\theta _W$ is the weak mixing angle. 
The CVC hypothesis requires $J_{\mu}^{\gamma (V)}=V_{\mu}^{3}$ and  the axial-vector coupling constant $g_A=-1.267$ is used. 
Iso-scalar component of the weak vector current $V_{\mu}^S$ was reported to be less than a few \% ~\cite{G0, HAPPEX} and is neglected. Iso-scalar component of the weak axial-vector current $A_{\mu}^S$ originates from the strange quark contribution and the first moment of the strange quark spin is set to $\Delta s=-0.08\pm0.02$~\cite{Hermes, COMPASS}. This iso-scalar component contributes to the NC cross section through $A_{\mu}^{3}+A_{\mu}^{S}$ of Eq.~(\ref{eq:a1}) and the neutrino cross section is larger by about 3\% than the anti-neutrino for the 12.97-MeV state. 
%textcolor{red}{Sakuda, 2022.0604: Suzuki-san, is this now ok?
%Suzuki, 2022.0707: It is OK.The vector mass $M_V$=0. GeVand the axial vector mass $M_A$=1.03 GeV.}

We define kinematic variables of the reaction as follows: $k^{\mu}$=($E$,$\vec{k}$) and $k^{ \prime \mu}$=($E^{\prime}$,$\vec{k}^{\prime}$) are the incident  and  scattered neutrino (or electron) four vectors, respectively, and $q^{\mu}=k^{\mu} -k^{\prime \mu}$=($\omega$, $\vec{q}$)=($E-E^{\prime}, \vec{k}-\vec{k}^{\prime}$) is a four momentum-transfer vector. We define $q\equiv |\vec{q}|$ and  $q_{\mu}^2 \equiv q^2-\omega^2$=$2EE^\prime(1-\cos\theta)$,  where   $\theta$ is the neutrino (or electron) scattering angle with respect to the incident particle direction and the electron mass is ignored. We note that $q^2=\omega^2 +|q_{\mu}^2| \ge | q_{\mu}^2|$.  In our application,  $J_i$ is the ground state (g.s.) of the $^{16}$O nucleus and  and $J_f$ is either $0^-$, $1^-$, $2^-$ or $1^+$.

We start with the NC(neutral-current) $^{16}$O($\nu, \nu ^{\prime}$) reaction.
The cross sections induced by $\nu$ or $\bar{\nu}$ are written in terms of  the multipole operators~\cite{Walecka75,Donnelly2,Donnelly79} as,
\begin{eqnarray}\label{diffcs}
 \displaystyle (\frac{d\sigma}{d\Omega})_{\frac{\nu}{\bar{\nu}^{\prime}}} &=&  \frac{2G_F^2 E_{\nu ^{\prime}}^2}{\pi} \frac{1}{2J_i+1}\cos^2\frac{\theta}{2} 
   \Biggl\{  \sum_{J=0}^{\infty} \big{|}\langle J_f \parallel \mathcal{M}_J(q)-\frac{\omega}{q}  \mathcal{L}_J(q)\parallel J_i\rangle\big{|}^2   \nonumber \\
    & &+ \Big{[}  \frac{|q_{\mu}^2|}{2q^2}+ \tan^2\frac{\theta}{2} \Big{]} \cdot \Big{[} \sum_{J=1}^{\infty} 
 \Big{(} \big{|}\langle J_f \parallel \mathcal{T}_{J}^{\rm el}(q) \parallel J_i \rangle \big{|}^2 + \big{|}\langle J_f \parallel \mathcal{T}_J^{\rm mag}(q) \parallel J_i\rangle\big{|}^2 \Big{)} \Big{]}  \nonumber \\
   & &\mp \tan\frac{\theta}{2} \sqrt{  \frac{|q_{\mu}^2|}{q^2}+ \tan^2\frac{\theta}{2} } \cdot \nonumber \\
    & & \Big{[} \sum_{J=1}^{\infty} 2 {\rm Re} \langle J_f \parallel \mathcal{T}_{J}^{\rm mag}(q) \parallel J_i \rangle \langle J_f \parallel \mathcal{T}_{J}^{\rm el}(q)\parallel J_i \rangle^{\ast} \Big{]} \Biggr\} ,
\label{eq:a2}
\end{eqnarray}
where $G_F$ is the Fermi constant, $\mathcal{M}_{J}(q)$, $\mathcal{L}_{J}(q)$, $\mathcal{T}_{J}^{\rm el}(q)$ and $\mathcal{T}_{J}^{\rm mag}(q)$ are the Coulomb, longitudinal, transverse electric and transverse magnetic multipole operators for the weak hadronic currents, respectively, and the transition matrix elements for these multipole operators are evaluated at the momentum transfer $q$. 
The minus (plus) sign ($\mp$)  corresponds to the neutrino (the anti-neutrino) case. 
The multipole operators are defined by the sum of the vector and axial-vector current operators for the neutral-current reactions,  
\begin{eqnarray}
\mathcal{M}_{J}(q) &=& M_{J}(q)+M_{J}^5(q),\  \mathcal{L}_{J}(q)=L_{J}(q)+L_{J}^5(q),\  \nonumber \\
\mathcal{T}_{J}^{\rm el}(q) &=&  T_{J}^{\rm el}(q)+T_{J}^{\rm el, 5}(q),\  {\rm and}\  \mathcal{T}_{J}^{\rm mag}(q)=T_{J}^{\rm mag}(q)+T_{J}^{\rm mag, 5}(q),
\end{eqnarray}
where the first term is due to the vector current and the second term with the index 5 is due to the axial-vector current. 
The concrete forms of the multipole operators and the transition matrix elements can be found in Ref.~\cite{OConnell, Walecka75, Donnelly79}. 

%% (Is it necessary to give the form factors? 
%% \begin{eqnarray}
%% \mathcal{F}_A&=&\frac{1}{2}\big(F_A^s\pm F_A\big)=\frac{1}{2}\frac{\Delta s\pm g_A}{(1- q^2/M^2_A)^2},\\
%% \mathcal{F}_p&=&\frac{2M^2\mathcal{F}_A}{m^2_\pi - q^2},
%% \end{eqnarray}
%% where the upper (lower) sign corresponds to proton (neutron) form factors, $\theta _W$ is the weak mixing angle, $m_\pi$ is the pion mass, $g_A=-1.2673$, and the strange quark contribution, the first moment of the strange quark spin, is set to $\Delta s=-0.08\pm0.02$~\cite{Hermes, COMPASS}. 

%The reduced matrix elements of these operators between the initial state $J_i$ and the final state $J_f$ like $\langle J_f %\parallel T_{J}^{\rm el} \parallel J_i \rangle$ are involved in the cross section calculations.  $G_{F}$ is the Fermi coupling %constant. 
The rate of the muon capture from the atomic $1s$ orbit of $^{16}$O is calculated as,   
\begin{eqnarray}\label{mucapture}
 \omega_{\mu} &=&  \frac{2G_F^2\cos^2\theta _C}{1+\nu/M_T}|\phi_{1s}|^2 \frac{1}{2J_i+1} \nonumber \\
    & &\cdot \Biggl\{\sum_{J=0}^{\infty} \big{|}\langle J_f \parallel \mathcal{M}_J(q)-\mathcal{L}_J(q) \parallel J_i\rangle\big{|}^2  
   + \sum_{J=1}^{\infty} 
 \big{|}\langle J_f \parallel \mathcal{T}_J^{\rm el}(q) - \mathcal{T}_{J}^{\rm mag}(q) \parallel J_i\rangle\big{|}^2  \Biggr\}.   
\end{eqnarray}
The multipole operators are evaluated at the momentum transfer $q=|\vec{q}|=|\vec{\nu}-\vec{k}|$, which is 95 MeV/c.

The differential $\beta^{-}$ decay rate $d\Lambda_{\beta^{-}}$ to yield an electron with energy $\epsilon$  in the solid angle $d\Omega_k$ and a neutrino in the solid angle $d\Omega_{\nu}$ is written as~\cite{Walecka75, Donnelly79},
\begin{eqnarray}\label{beta-decay}
d\Lambda_{\beta^{-}} &=&  \frac{G_F^2 \cos^2\theta_c}{2\pi ^2}k\epsilon (W_0-\epsilon)^2 d\epsilon  \frac{d\Omega_k}{4\pi}\frac{d\Omega_{\nu}}{4\pi} \frac{4\pi}{2J_i+1}
   \Biggl\{  \sum_{J=0}^{\infty} \big{|}\langle J_f \parallel \mathcal{M}_J(q)-\mathcal{L}_J(q)\parallel J_i\rangle\big{|}^2   \nonumber \\
    & &+ \Big{[} \sum_{J=1}^{\infty} 
\big{|}\langle J_f \parallel \mathcal{T}_{J}^{\rm mag}(q) + \mathcal{T}_J^{\rm el}(q) \parallel J_i\rangle\big{|}^2  \Big{]}  \Biggr\} ,
\end{eqnarray}
where  $\theta _C$ is the Cabibbo angle and $W_0$ is the maximum value of $\epsilon$. The multipole operators are evaluated at the momentum transfer $q=|\vec{q}|=|\vec{k}+\vec{{\nu}}|$ ($0\le q \le$10.4 MeV/c). 

We now describe the electromagnetic ($e,e^{\prime}$) reaction.  
When the electron is scattered by the nucleus with the momentum transfer $q$ from the ground state ($J_i=0^+$) to the excited states ($J^P$), the differential ($e,e^{\prime}$) cross section $(\frac{d\sigma}{d\Omega})_{(e,e^{\prime})}$ and the form factors 
are written in the following form~\cite{deForest,Spamer,Deutschmann, Donnelly2}\footnote{We write the definition of the form factors as  in Ref.~\cite{Sick, Donnelly1},  which is different from others~\cite{Stroetzel,  Kim} by a  factor $ 4\pi /Z^2 $.  A special care should be taken when the data are plotted in the same figure.},
%Note2FF},
\begin{eqnarray}\label{magnetic}
(\frac{d\sigma}{d\Omega})_{e,e^{\prime}} &=& 4\pi \sigma_{\rm Mott} F^2(q)/R_{\rm recoil},     \\    
\sigma_{\rm Mott} &=& \Big{[} \frac{\alpha \cos\frac{\theta}{2}}{2E \sin ^2\frac{\theta}{2} } \Big{]}^2,     \\  
 F^2(q)  &=& (\frac{|q_{\mu}^2|}{q^2})^2 F_L^2(q)+(\frac{|q_{\mu}^2|}{2q^2}+\tan^2\frac{\theta}{2})F_T^2(q), \\  
  F_L^2(q)  &=& \frac{1}{2J_i+1}\sum_{J=0}^{\infty} \big{|}\langle J \parallel \tilde{M}_J(q) \parallel J_i\rangle\big{|}^2,  \\ 
 F_T^2(q)  &=& \frac{1}{2J_i+1}\sum_{J=1}^{\infty}\{ \big{|}\langle J \parallel \tilde{T}_J^{\rm el}(q) \parallel J_i\rangle\big{|}^2 + \big{|}\langle J \parallel \tilde{T}_J^{\rm mag}(q) \parallel J_i\rangle\big{|}^2 \},    \\
R_{\rm recoil} &=& 1+(2E \sin^2\frac{\theta}{2})/{M_T},
\end{eqnarray}  
where $\sigma_{\rm Mott}$ is the Mott cross section, $F^2(q)$ is the nuclear form factor, $F_L^2(q)$ and $F_T^2(q)$ are the longitudinal and transverse form factors, respectively,  $\tilde{M}_{J}(q)$,  $\tilde{T}_{J}^{\rm el}(q)$, and $\tilde{T}_J^{\rm mag}(q)$  are the Coulomb, electric and magnetic multipole ($J$) operators for the hadronic current induced by the electromagnetic interaction, respectively, and $M_T$ is the mass of the target nucleus, and $R_{\rm recoil}$ is the recoil correction. 
We assign tilde ($\tilde{\ }$) on the electromagnetic multipole operators and they have both the iso-vector and iso-scalar components. The CVC requires that the iso-vector component of $\tilde{M}_{J}(q)$,  $\tilde{T}_{J}^{\rm el}(q)$, and $\tilde{T}_J^{\rm mag}(q)$ is equal to $M_{J}(q)$,  $T_{J}^{\rm el}(q)$, and $T_J^{\rm mag}(q)$, respectively. The sum over $J$ may be just one dominant multipole ($J$) or it is taken over some multipoles ($J$) being excited during the interaction. In our case,  $J_i = 0^+_{\rm g.s.}$  and $J_f$ is either  $0^-, 1^-, 2^-,$  or $3^- $.  We use the reduced transition probability $B(M2, q)$ to discuss the isospin mixing parameter in the present paper. The quantity $B(E_J, q)$ or $B(M_J, q)$ is sometimes used instead of the form factors, since it is directly related to the electric or magnetic multipole  moment~\cite{ABohr, Spamer, OConnell} which is measured in an experiment. The reduced magnetic transition probability $B(M_J, q)$ is related to the magnetic form factors $F_T^2(M_J,q)$ as~\cite{Spamer, Deutschmann}, 
\begin{eqnarray}
B(M_J, q)=\frac{J[(2J+1)!!]^2}{J+1}q^{-2J}F_T^2(M_J, q).
\label{eq:a12}
\end{eqnarray}  
Stroetzel measured $\sqrt{B(M2, q)}$~\cite{Stroetzel}, while others measured $F_T^2(M2, q)$~\cite{Kim, Sick}. We converted  the values of $F_T^2(M2, q)$ of Kim ${\it et \ al.}$~\cite{Kim}  to $\sqrt{B(M2, q)}$ using the above equation. No Coulomb corrections were applied, which may be significant at the low incident energy below 50 MeV~\cite{Deutschmann}. 

\section{The quenching factor $f_s=g_s^{\rm eff}/g_s$ of the spin $g$ factor $g_s$ and the quenching factor $f_A=g_A^{\rm eff}/g_A$ of the weak axial-vector coupling constant $g_A$}

Donnelly and Walecka of Ref.~\cite{Donnelly1,Donnelly2,Walecka75} analysed the data of $^{16}$O$(e, e^{\prime})^{16}$O$(E_x$=12-20 MeV) scattering and semi-leptonic weak interactions (muon capture and $\beta$ decay) and obtained the reduction factors ($a/\xi$=0.6-0.7) to the transition amplitudes of their model to calculate the neutrino-$^{16}$O cross sections at $E_x$=13-19 MeV  precisely with accuracy of 15-20\%. This reduction in transition amplitudes of a calculation model (or in the coupling constant) is sometimes called a quenching factor. At the time of this analysis, the isospin mixing of the two 2$^-$ states at 12.53 MeV and 12.97 MeV was not known and was not considered. We follow this analysis and determine the correction factors (quenching factors)  of both the axial-vector coupling constant and the electromagnetic spin $g$ factor by considering the isospin mixing effect in ($e,e^{\prime}$) data and using newer data of the muon capture and $\beta$ decay.  

First, we determine the quenching factor $f_s=g_s^{\rm eff}/g_s$ of the spin $g$ factor $g_s$ for the magnetic form factors for the excited states of $^{16}$O at $E_x$=12-13 MeV in order to get a reliable calculation of the ($e,e^{\prime}$) cross section. 
%Though a quenching factor of  $f_s$=0.65 and $f_s= 0.70\pm 0.05$ was found in the ($e, e'$) cross section for $^{12}$C(2$^-, T = 1$, 19.40 MeV) state using the Cohen-Kurath Hamiltonian~\cite{CohenKurath,Gaade} and using the SFO shell model~\cite{SFO}, respectively, 
The evaluation of the quenching factor $f_s$ must be performed for $^{16}$O($E_x$=12-13 MeV, $T$ = 1) using ($e, e^{\prime}$) cross section. Thus, we use the data of $^{16}$O($E_x$=13.09 MeV, 13.25 MeV, $T$ = 1), but we do not use those of $^{16}$O(12.97 MeV and 12.53 MeV, 2$^-$), since the electromagnetic interaction mixes the iso-vector ($T = 1$)  and iso-scalar ($T = 0$) states according to Eq.~(\ref{eq:a1}) and the data may have been affected by a possible isospin mixing effect. 

 Fig.~\ref{FM2} shows the data (squares, triangles and circles) and the predictions (lines) of the transverse form factor $F^2_T(q)$ of the ($e, e^{\prime}$) cross section near 13 MeV~\cite{Stroetzel, Kim, Sick}. We find that a quenching factor $f_s= 0.65\pm 0.05$ reproduces the data for 13.09 MeV (1$^-$) and 13.25 MeV (3$^-$) states well. We do not show the older data of Ref.~\cite{Vanpraet}, since their statistics is very low and their data are inconsistent with those shown in the figure. 
 We note that we use data of $F^2_T(q)$ for only $q>$1.5 fm$^{-1}$  in this evaluation to exclude a possible isospin mixing effect of 12.97 MeV and 12.53 MeV. Thus, we use  $f_s$= 0.65 to evaluate the isospin mixing between the 12.53 MeV and 12.97 MeV in Section~\ref{sec3}. 
 
Next, we also determine the quenching factor $f_A=g_A^{\rm eff}/g_A$ of the weak axial-vector coupling constant $g_A$ in order to calculate the $^{16}$O($\nu, \nu^{\prime}$) cross section for the 12.97 MeV state precisely. In this calculation,  we use a quenching factor $f_s$= 0.65 for the spin $g$ factor in the magnetic form factors. 

The SFO-tls model was found to reproduce the experimental data of the muon capture on $^{16}$O~\cite{muOcapture} within 10\%, with a quenching factor $f_A=g_A^{\rm eff}/g_A$=0.95~\cite{SFOtls}. But, the total muon capture rate was mainly determined by the contribution of the giant resonances at $E_x>$16 MeV, since the momentum transfer $q=E_{\nu}$ is about 95 MeV/c.  
 The evaluation of the quenching factor $f_A$ must be carried out for $^{16}$O(12.97 MeV, 2$^-, T = 1$). This time, the CC reactions such as the rate of the muon  capture on $^{16}$O(g.s.) to  $^{16}$N~\cite{Cohen1, Cohen2, Deutsch1, Kane, Eckhause}  and the $\beta^-$ decay from the ground state ($2^-$, $T=1$) of $^{16}$N to the ground state ($0^+$)  of $^{16}$O~\cite{Ajzenberg86, Warburton, Heath} can be used, since the CC reactions are induced only by the iso-vector ($T=1$) weak vector and axial-vector current $J_{\mu}^{\rm CC}$ and the iso-scalar ($T=0$) component is zero. The ground state of $^{16}$N(0 MeV, 2$^-$, $T=1$, $T_3=-1$) is the isobaric analog state of the excited state $^{16}$O(12.97 MeV, 2$^-$, $T=1$, $T_3=0$) and the spacial part of their wave functions is common.   
 The comparison of the experimental data with the shell model using various $f_A$ is given in Table~\ref{MuonCapture} and we find that while $f_A$=0.62$\pm$0.03 reproduces the data of the muon capture rate, $f_A$=0.73$\pm$0.02 reproduces the $\beta^-$ decay rate reasonably well. We note that the transition amplitudes for the muon capture are evaluated at $q$=95 MeV/c and that those for the  $\beta^-$ decay rate are evaluated at 0$<q<$10 MeV/c. Taking a simple mean and the variation of the two values, we use $f_A$=0.68$\pm$0.05 in the calculation of the neutrino-$^{16}$O(12.97  MeV, 2$^-$) cross sections. When we calculate the neutrino-$^{16}$O($E_x>$16 MeV) cross sections, we use $f_A$=0.95 as in Table~\ref{MuonCapture} and Ref.~\cite{SFOtls}.

%%\hline
%%\hline
%%\end{tabular}

\begin{table*}[!]
\caption{Rate of the partial muon capture ($\mu^-, \nu_{\mu}$) from the 1s orbit on $^{16}$O(g.s., 0$^+$) to the bound states (2$^-$(ground state), 0$^-$, 3$^-$, 1$^-$, $T=1$) of $^{16}$N and the total muon capture rate from $^{16}$O to $^{16}$N(g.s., 2$^-$), in units of 10$^3$ 1/s. The $\beta^-$-decay rate from the ground state of $^{16}$N to $^{16}$O(g.s.) is also shown. The energy $E_x$ is given with respect to the ground state (2$^-$) of $^{16}$N.}
\begin{tabular}{>{\centering \arraybackslash}p{4.0cm}>{\centering \arraybackslash}p{3.0cm}>{\centering \arraybackslash}p{3.0cm}>{\centering \arraybackslash}p{4.0cm}}
\hline
\hline
 Weak Process & States of $^{16}$N & Experimental &  Model Prediction~\cite{SFOtls}      \\
 & $E_x$ MeV($J^P$) &   Data [Reference]     & (with $f_A=g_A^{\rm eff}/g_A$)\\
 &  &   &     \\  
 \hline
 $\mu$ capture  (10$^3$ 1/s)& 0 MeV(2$^-$) & 6.3$\pm$0.7~\cite{Cohen1,Cohen2}  & 7.2    \\
     &   & 7.9$\pm$0.8~\cite{Deutsch1}  &  ($f_A$=0.63$\pm$0.03)   \\
     &  & 8.0$\pm$1.2~\cite{Kane}  &     \\
   & 0.120 MeV(0$^-$) & 1.1$\pm$0.2~\cite{Cohen1,Cohen2}   & 1.33  \\
   &    & 1.56$\pm$0.18~\cite{Kane}  &  ($f_A$=0.62$\pm$0.02)  \\
    & 0.298 MeV(3$^-$) & $<$0.09~\cite{Kane}   & $f_A<$0.60  \\
  & 0.397 MeV(1$^-$) & 1.73$\pm$0.10~\cite{Cohen1,Cohen2}   & 1.52    \\
   &    & 1.31$\pm$0.11~\cite{Kane}  &   ($f_A$=0.62$\pm$0.03) \\
 &  Sum(2$^-$+1$^-$+0$^-$) & 9.15$\pm$0.70~\cite{Cohen1,Cohen2}   & 10.1$\pm$0.5   \\
    &  & 10.9$\pm$0.7~\cite{Deutsch1}   &   ($f_A$=0.62$\pm$0.02)  \\
   &  & 10.87$\pm$1.22~\cite{Kane}   &     \\
 & $E_x>$5 MeV  & 102.6$\pm$0.6~\cite{muOcapture}     & 112.0 \\
    &  & (0.98$\pm$0.03)$\times$10$^2$~\cite{Eckhause}  &  ($f_A$=0.95)     \\
\hline
\hline
$^{16}$N $\beta^-$ decay rate &  & &  \\
 $\Lambda_{\beta^-}$ ($\times$10$^{-3}$ 1/s) & 2$^- \to 0^+$ & 27.2$\pm$0.4~\cite{Alburger,Warburton,Heath, Ajzenberg86}   &    27.2  ($f_A$=0.73$\pm$0.01)\\
%%  $\Lambda_{\beta}$ (1/s) & 0$^- \to 0^+$ & 0.489$\pm$0.020~\cite{Heath}   &   \\
\hline
\end{tabular}
\label{MuonCapture}
\end{table*} 

\begin{figure}[ht!]
\centering
\includegraphics[width=14.0cm,scale=1.0]{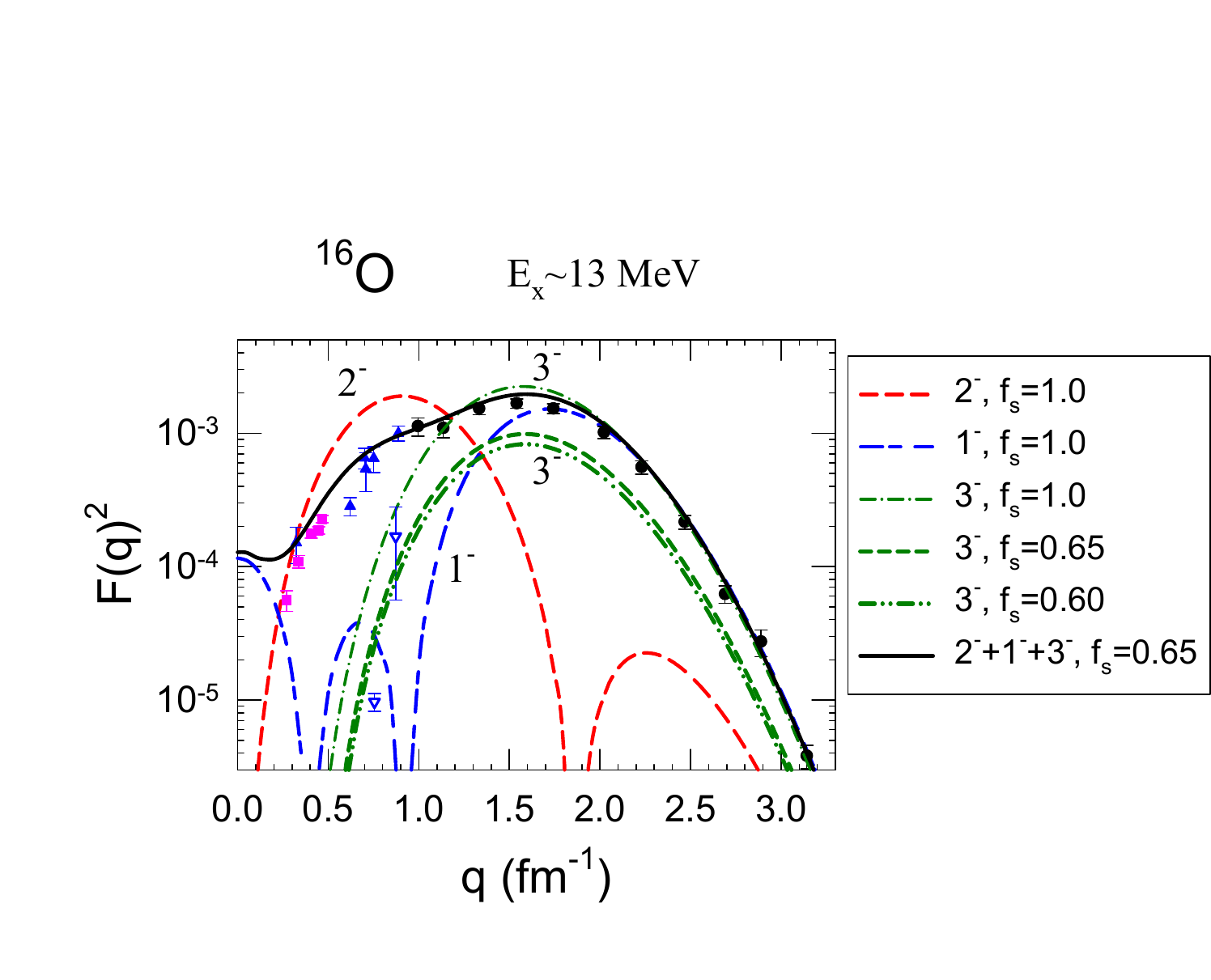}
\caption{Transverse form factor $F^2_T(q)$ of the inelastic electron scattering as a function of the momentum transfer  $q\ ({\rm fm}^{-1})$ for the multipoles near 13 MeV including 12.97 MeV (2$^-$), 13.09 MeV (1$^-$) and 13.26 MeV (3$^-$) states. 
Data are taken from Stroetzel (red squares, Ref.~\cite{Stroetzel}), Kim (blue triangles, Ref.~\cite{Kim}) and Sick (black circles, Ref.~\cite{Sick}). The predictions of
the M2, E1 and M3 form factors with $f_s$ = $g_s^{\rm eff}$/$g_s$ =1.0 are shown by red dashed, blue long-short dashed and green dash-dotted lines, respectively. Black solid line denotes the prediction of the sum of the M2, E1 and M3 form factors with $f_s$ =0.65. 
The M3 form factors with $f_s$ = 0.65 and 0.60 are shown by green dashed and dashed-two-dotted lines, respectively. In this figure, no isospin mixing is assumed ($\beta$=0).
\label{FM2}}
\end{figure}

}  %%%%%% Appendix end



\nocite{*}

\begin{thebibliography}{99}
\bibitem{Karmen1} B. Bodman {\it et al.} (KARMEN Collaboration), Phys. Lett. B \textbf{267}, 321 (1991).
\bibitem{Karmen2} B. Armbruster {\it et al.} (KARMEN Collaboration), Phys. Lett. B \textbf{423}, 15 (1998).
\bibitem{Coherent} D. Akimov {\it et al.} (COHERENT Collaboration), Science {\bf 357}, 1123 (2017).
\bibitem{Kolbe} E. Kolbe, K. Langanke, S. Krewald, and F.-K. Thielemann, Nucl. Phys. A \textbf{540}, 599 (1992).
\bibitem{Kolbe1}  K. Langanke, P. Vogel and E. Kolbe, Phys. Rev. Lett. \textbf{76}, 2629 (1996).
\bibitem{Beacom1} J. Beacom and P. Vogel, Phys. Rev. D \textbf{58}, 053010 (1998).
\bibitem{Kolbe02}
E. Kolbe, K. Langanke, and P. Vogel, Phys. Rev. D {\bf 66}, 013007 (2002).
\bibitem{SNrev1} H.-Th. Janka, K. Langanke, A. Marek, G. Martinez-Pinedo, and B. Mueller,  Phys.~Rept.\textbf{442}, 38(2007).
\bibitem{SNrev2} A. Mirizzi, I. Tamborra, H.-T. Janka, N. Saviano, K. Scholberg, R. Bollig, L. H$\ddot{\mbox{u}}$depohl, and S. Chakraborty, Riv. Nuovo Cim. 39, 1 (2016).
\bibitem{SNrev3} 
S. Horiuchi and J. P.  Kneller,  J. Phys. G: Nucl.~Part.~Phys. \textbf{45}, 043002 (2018).
\bibitem{SNrev4} E. Vitagliano, I. Tamborra, and G. Raffelt, Rev. Mod. Phys. {\bf 92}, 045006 (2020).
\bibitem{SK-SN1} M. Ikeda {\it et al.} (Super-K Collaboration), Astrophys. J. \textbf{669}, 519 (2007).
\bibitem{SK-SN2} K. Abe {\it et al.} (Super-K Collaboration), Astropart. Phys. \textbf{81}, 39 (2016).
\bibitem{Haxton87}
W. C. Haxton, Phys. Rev. D {\bf 36}, 2283 (1987).
\bibitem{Kolbe03}
E. Kolbe, K. Langanke, G. Mart\'inez-Pinedo, and P. Vogel, J. Phys. G {\bf 29}, 2569 (2003).
\bibitem{Nakazato1} K. Nakazato, T. Suzuki and M. Sakuda, PTEP \textbf{2018}, 123E02 (2018).
\bibitem{SKsolar1}
Y. Fukuda  {\it et al.} (Super-K Collaboration), Phys. Rev. Lett.\textbf{81}, 1158 (1998).
\bibitem{SKsolar2}
S. Fukuda  {\it et al.} (Super-K Collaboration), Phys. Rev. Lett.\textbf{86}, 5651 (2001).
\bibitem{SKsolar3} S. Fukuda  {\it et al.} (Super-K Collaboration), Phys. Rev. Lett.\textbf{86}, 5656 (2001).
\bibitem{SK-det}
K. Abe {\it et al.} (Super-K Collaboration), Phys. Rev. D\textbf{94}, 052010 (2016).    
\bibitem{EGADS}
LI.  Marti {\it et\  al.} (Super-K Collaboration),  Nucl. Instrum. Meth. A \textbf{959}, 163549 (2020).
\bibitem{SK-Gd1}
K. Abe {\it et\ al.} (Super-K Collaboration), Nucl. Instrum. Meth. A \textbf{1027}, 166248 (2022).
\bibitem{Vagins}
 J. F. Beacom and M. R. Vagins, Phys. Rev. Lett. \textbf{93}, 171101 (2004).
\bibitem{SK-SRN}
K. Abe {\it et\ al.} (Super-K Collaboration), Phys. Rev. D\textbf{104}, 122002 (2021). 
\bibitem{KamLAND-SRN}
S. Abe {\it et\ al.} (KamLAND Collaboration), Astrophys.J. \textbf{925}:14, 1 (2022).   
%%
%%% Chapter II 
%%
\bibitem{Adelberger} E. Adelberger, R. Marrs, K. Snover and J. Bussoletti, Phys. Rev. C{\bf 15}, 484 (1977).
\bibitem{Flanz} J.B. Flanz, R. Hicks, R. Lindgren, G. Peterson, J. Dubach and W. Haxton, Phys. Rev. Lett. {\bf 43}, 1922 (1979).
\bibitem{Cosel} P. von Neumann-Cosel,  H.-D. Gr\"af,  U. Kr\"amer,  A. Richter and E. Spamer,  Nucl.Phys. A\textbf{669},  3 (2000).
\bibitem{Auerbach1}
N. Auerbach, Phys. Rep. \textbf{98}, 273 (1983).
\bibitem{Auerbach2}
B. M. Loc, N. Auerbach and G. Colo,  Phys. Rev. C\textbf{99}, 014311 (2019).
\bibitem{Stroetzel}   M. Stroetzel, Z. f. Physik \textbf{214},  357 (1968).
\bibitem{Wagner}  G.J. Wagner, K.T. Knopfle, G. Mairle, F. Doll, H. Hafner and J.L.C. Ford, Jr., Phys. Rev. C \textbf{16},  1271 (1977).
\bibitem{Leavitt}  R. A. Leavitt, H. C. Evans, G. T. Ewan, H.-B. Mak, R. E. Azuma, C. Rolfs, and K. P. Jackson, Nucl. Phys. A \textbf{410}, 93 (1983).
\bibitem{Zijderhand}  F. Zijderhand and C. van der Leun, Nucl. Phys. A \textbf{460}, 181 (1986).
\bibitem{Charity}  R.J. Charity {\it et al.}, Phys. Rev. C \textbf{99}, 044304 (2019).
\bibitem{ENSDF} The Evaluated Nuclear Structure File (ENSDF), http://www.nndc.bnl.gov/ensdf/. 
\bibitem{Tilley}
D.R. Tilley, H.R. Weller and C.M. Cheves, Nucl. Phys. A \textbf{565}, 1 (1993).    
%%
%%% Chapter III
%%
\bibitem{Kim}  J.C.  Kim, R.P.  Singhal and H.S.  Caplan, Can.  J. Phys.\textbf{48},  83 (1970).
\bibitem{Vanpraet}  G.J. Vanpraet and W.C. Barber,  Nucl.Phys. \textbf{79},  550 (1966).
\bibitem{Sick}  I. Sick, E.B. Hughes, T.W. Donnely, J.D. Walecka and G.E. Walker, Phys. Rev.Lett. \textbf{23},  1117 (1969).
\bibitem{Donnelly1}   T.W. Donnelly and J.D. Walecka, Phys.Lett. B\textbf{41},   275 (1972).
\bibitem{Donnelly2}   T.W. Donnelly and J.D. Walecka,  Ann.Rev.Nucl.Sci.\textbf{25},  329 (1975).
\bibitem{Walecka75}
J. D. Walecka, in {\it Muon Physics}, eds V. H. Hughes and C. S. Wu (Academic, New York 1975), Section~4,  Page 113-217.
%%
%%% Chapter IV
%%
\bibitem{SFOtls}
T. Suzuki, and T. Otsuka, Phys. Rev. C {\bf 78}, 061301(R) (2008).
\bibitem{suzuki11}
T. Suzuki, J. Phys.: Conf. Series {\bf 321}, 012041 (2011).
\bibitem{YSOX} C. Yuan, T. Suzuki, T. Otsuka, F. Xu, and N. Tsunoda,  Phys. Rev. C \textbf{85}, 064324 (2012).
\bibitem{SFO} T. Suzuki, R. Fujimoto, and T. Otsuka, Phys. Rev. C \textbf{67}, 044302 (2003).
\bibitem{suzuki06}
T. Suzuki, S. Chiba, T. Yoshida, T. Kajino, and T. Otsuka, Phys. Rev. C \textbf{74}, 034307 (2006).
\bibitem{TSuzuki2} T. Suzuki, S. Chiba, T. Yoshida, K. Takahashi, and H. Umeda,  Phys. Rev. C \textbf{98}, 034613 (2018).
\bibitem{Jachowicz} N. Jachowicz, S. Rombouts, K. Heyde, and J. Ryckebusch, Phys. Lett. B \textbf{564}, 42 (2003).
\bibitem{OConnell} J.S. O'Connell, T. Donnelly and J. Walecka, Phys. Rev. C. \textbf{6}, 719 (1972).
\bibitem{Strumia} A. Strumia and S. Vissani, Phys. Lett. B \textbf{564}, 42-54 (2003).
%\bibitem{Note3} We did not use the simple analytic form $\sigma(E_{\nu})=(0.75\times 10^{-47}\ {\rm cm}^2)(E-15)^4$,  which is given in Ref.~\cite{Beacom1}, since it disagrees with the cross section ($\times 0.3$) given by Ref.~\cite{Kolbe02}.
\bibitem{Kawabata} T. Kawabata {\it et\ al.},  Phys. Rev. C\textbf65,  064316 (2002).
\bibitem{Kuchler}  G. Kuchler, A. Richter, E. Spamer, W. Steffen and W. Knupfer,  Nucl. Phys. A\textbf{406},  473 (1983).
\bibitem{Wright}  C.E. Hyde-Wright {\it et\ al.},  Phys. Rev. C\textbf{35},  880 (1987).
\bibitem{Reen} M.S. Reen {\it et\ al.},  JPS Conf. Proc.\textbf{31}, 011054 (2020).
%%
%%% Chapter V
%%
\bibitem{OConnor} E. O'Connor {\it et al.}, J. Phys. G: Nucl. Part. Phys. \textbf{45}, 104001 (2018).
\bibitem{Just} O. Just, R. Bollig, H.-Th. Janka, M. Obergaulinger, R. Glas and S. Nagataki, Mon. Not. Roy. Astron. Soc. \textbf{481}, 4786 (2018).
\bibitem{Richers} S. Richers, H. Nagakura, C. D. Ott, J. Dolence, K. Sumiyoshi and S. Yamada, Astrophys. J. \textbf{847}, 133 (2017).
\bibitem{Nagakura} H. Nagakura, A. Burrows, and D. Vartanyan, MNRAS \textbf{506}, 1462 (2021). 
\bibitem{Keil} M. Th. Keil, G. Raffelt and H.-Th. Janka, Astrophys. J. \textbf{590}, 971 (2003).
\bibitem{Tamborra} I. Tamborra, B. M\"{u}ller, L. H\"{u}depohl, H.-Th. Janka and G. Raffelt, Phys. Rev. D\textbf{86}, 125031 (2012).
\bibitem{Nakazato} K. Nakazato, K. Sumiyoshi, H. Suzuki, T. Totani, H. Umeda and S. Yamada, Astrophys. J. Suppl. \textbf{205}, 2 (2013).
\bibitem{Jost} 
K. Abe {\it et al.} (Hyper-K Collaboration), Astrophys. J. \textbf{916}, 15 (2021).
\bibitem{Artur} 
A.Ankowski, O.Benhar, T.Mori, R. Yamaguchi and M.Sakuda, Phys.Rev.Lett.{\bf 108}, 052505 (2012). 
\bibitem{T2K6MeV1}
K. Abe {\it et\ al.} (T2K Collaboration), Phys. Rev. D\textbf{90}, 072012 (2014).
\bibitem{T2K6MeV2} K. Abe {\it et\ al.} (T2K Collaboration), Phys. Rev. D\textbf{100}, 112009 (2019).
%%
%%% Chapter VI
%%
\bibitem{Cosel2016} 
Peter von Neumann-Cosel, 180$^\circ$ Electron Scattering at the S-DALINAC, JPS Conf. Proc. \textbf{12}, 010030 (2016). 
\bibitem{MESA}B.S. Schlimme {\it et al.} (MAGIX Collaboration),  Nucl.Instrum.Meth. A \textbf{1013}, 165668 (2021). 
\bibitem{Suda1} 
T. Suda, Y. Honda, Y. Maeda and K. Tsukada, Proton Radius, KEK News Vol.\textbf{40}, No.3, 107 (2021)(in Japanese). 
\bibitem{Suda2} T. Suda {\it et al.}, J. Particle Accelerator Society Japan \textbf{15}, 52
(2018) (in Japanese). 
%%
%%% Appendix A
%%
\bibitem{KarmenCC1}
B. Bodmann {\it et al.}(KARMEN Collaboration), Phys. Lett. B \textbf{339}, 215 (1994).
\bibitem{KarmenCC2} B. Bodmann {\it et al.}(KARMEN Collaboration), Phys. Lett. B \textbf{332}, 251 (1994).
%\bibitem{KarmenNC1}
%B. Bodmann {\it et al.} (KARMEN Collaboration), Phys. Lett. B 267, 321 (1991).
%\bibitem{KarmenNC2} B. Armbruster{\it et al.}(KARMEN Collaboration), Phys. Lett. B 423, 15 (1998).
\bibitem{LSNDCCe}
  L.B. Auerbach {\it et al.} (LSND Collaboration), Phys. Rev. C \textbf{64}, 065501 (2001). 
\bibitem{LSNDCCm1}
  C. Athanassopoulos {\it et al.} (LSND Collaboration), Phys. Rev. C \textbf{56}, 2806 (1997).
\bibitem{LSNDCCm2}  L.B. Auerbach {\it et al.} (LSND Collaboration), Phys. Rev. C \textbf{66}, 015501 (2002). 
\bibitem{VMU} 
T. Otsuka, T. Suzuki, M. Honma, Y. Utsuno, N. Tsunoda, K. Tsukiyama, and M. Hjorth-Jensen, Phys. Rev. Lett. 104, 012501 (2010).
\bibitem{WB1992} E. K.Warburton and B. A. Brown, Phys. Rev. C 46, 923 (1992).
\bibitem{HAPPEX} Z. Ahmed {\it et al.} (HAPPEX Collaboration),  Phys.Rev.Lett.\textbf{108}, 102001 (2012).
\bibitem{G0} D. Androic {\it et al.} (G0 Collaboration),  Phys.Rev.Lett.\textbf{104}, 012001 (2010).
\bibitem{COMPASS} V. Yu. Alexakhin {\it et al.} (COMPASS Collaboration),  Phys.Lett.B\textbf{647}, 8 (2007).
\bibitem{Hermes} A. Airapetian {\it et al.} (Hermes Collaboration),  Phys.Rev.D\textbf{75}, 012007 (2007).
\bibitem{Donnelly79}   T.W. Donnelly and R.D. Peccei,  Phys.Rept. \textbf{50},   1 (1979).
\bibitem{deForest}   T.  de Forest, Jr.  and J.D. Walecka,  Adv.Phys.\textbf{15},   1 (1966).
\bibitem{Spamer} E. Spamer, Z.Phys.   \textbf{191}, 24 (1966).
\bibitem{Deutschmann} U. Deutschmann, G. Lahm, R. Neuhausen and J.C. Bergstrom, Nucl. Phys. A\textbf{411}, 337 (1983).
\bibitem{ABohr}
A. Bohr and B.T. Mottelson, Nuclear Structure Volume 1, World Scientific Publishing Co.
%%
%%% Appendix B
%%
\bibitem{muOcapture}  T. Suzuki,  D.F. Measday and J.P. Roalsvig, Phys. Rev. C \textbf{35},  2212 (1987).
\bibitem{Cohen1} R.C. Cohen, S. Devons and A. Kanaris, Phys.Rev.Lett.\textbf{11}, 134(1963).
\bibitem{Cohen2} R.C. Cohen, S. Devons and A. Kanaris, Nucl.Phys.\textbf{57}, 255(1964).
\bibitem{Deutsch1} J. Deutsch, L. Grenacs, P. Igo-Kemenes, P. Lipnik and P. Macq, IL Nuovo Cimento B\textbf{52}, 557 (1967).
\bibitem{Kane} R. Kane, M. Eckhause, G. Miller, B. Roberts, M. Vislay and R. Welsh, Phys.Lett.B\textbf{45}, 292(1973).
\bibitem{Eckhause} M. Eckhause, T. Filippas, R. Sutton and R. Welsh,  Phys. Rev. \textbf{132}, 422 (1963).
\bibitem{Ajzenberg86} F. Ajzenberg-Selove, Nucl. Phys. A\textbf{460}, 1 (1986).
\bibitem{Warburton} E. Warburton, D. Alburger and D. Millener, Phys. Rev. C\textbf{29}, 2281 (1984).
\bibitem{Heath} A. Heath and G. Garvey, Phys. Rev. C\textbf{31}, 2190 (1985).
\bibitem{Alburger} D. Alburger, A. Gallmann and D. Wilkinson, Phys. Rev. \textbf{116}, 939 (1959).

\end{thebibliography}
\end{document}